\documentclass[aip,jcp,amsmath,amssymb,preprint]{revtex4-2}

\usepackage{graphicx}
\usepackage{dcolumn}
\usepackage{bm}

\usepackage{xcolor}
\usepackage[utf8]{inputenc}
\usepackage[T1]{fontenc}
\usepackage{mathptmx}
\usepackage{etoolbox}
\usepackage{dsfont}

\renewcommand{\Re}{\operatorname{Re}}

\makeatletter
\def\@email#1#2{%
 \endgroup
 \patchcmd{\titleblock@produce}
  {\frontmatter@RRAPformat}
  {\frontmatter@RRAPformat{\produce@RRAP{*#1\href{mailto:#2}{#2}}}\frontmatter@RRAPformat}
  {}{}
}%
\makeatother
\begin{document}


\title{Collective intermolecular Coulombic decay beyond Coulomb}
\author{Alan G. Falkowski}
\email{alangf@unicamp.br}
\affiliation{Theoretische Chemie, Physikalisch-Chemisches Institut, Universit\"at Heidelberg, Im Neuenheimer Feld 229, Heidelberg D-69120, Germany}
\affiliation{Instituto de Física “Gleb Wataghin”, Universidade Estadual de Campinas, 13083-859 Campinas, São Paulo, Brazil}
\author{Lorenz S. Cederbaum}
\email{Lorenz.Cederbaum@pci.uni-heidelberg.de}
\affiliation{Theoretische Chemie, Physikalisch-Chemisches Institut, Universit\"at Heidelberg, Im Neuenheimer Feld 229, Heidelberg D-69120, Germany}

\date{\today}

\begin{abstract}
Intermolecular Coulombic decay (ICD) is a widely spread phenomenon in nature and laboratory in which the excess energy of a donor is utilized to ionize a nearby acceptor. If the excess energy is insufficiently large to enable ICD, two (or more) donors can collectively transfer their combined excess energy to ionize the acceptor. Experiments show that this collective ICD is, surprisingly,  operative in gases. Recently, it has been demonstrated that standard ICD can efficiently take place at large distances between the donors and acceptors due to retardation. Here, we derive the theory of collective ICD including retardation. Quantum electrodynamics (QED) perturbation theory is used and it is shown that the theory can be substantially simplified and the process also made more amenable to interpretation by introducing two-body interaction potentials which include retardation. Explicit formulas for the rate of collective ICD are derived and interpreted by expressing the rate in terms of measurable quantities and geometric factors. It is demonstrated that the change of the permanent dipole moments of the species upon excitation is a relevant ingredient in collective ICD.
\end{abstract}

\maketitle

\section{Introduction}

In the interatomic and intermolecular Coulombic decay (ICD) process, an initially excited donor transfers its excess energy to a neighboring acceptor, ionizing the latter when the transferred energy exceeds its ionization threshold \cite{cederbaum1997giant}.  
Since its theoretical prediction \cite{cederbaum1997giant} and first experimental confirmation \cite{marburger2003experimental,jahnke2004experimental}, ICD has proven to be an efficient channel for producing ionized species in weakly bound systems, such as clusters, liquids, and solvated ions. See the comprehensive reviews in \cite{hergenhahn2011interatomic,jahnke2015interatomic,jahnke2020review}.

In these environments, the asymptotic ICD rate (for bright states) of two electronic species separated by a distance $R$ is proportional to $1/R^6$ due to the bare Coulomb interaction between species ($1/R^3$) \cite{santra2002nonhermitian}.
It is well established that retardation effects introduce extra components into the ICD rate, adding an intermediate $1/R^4$ term and a long-range $1/R^2$ term \cite{hemmerich2018influence,cederbaum2025relativistic}. The impact of retardation in the studied systems is, however, minor.

Very recently, we showed that ICD can be efficiently operative in low-density gases, where atoms and molecules are typically separated by micrometers. It has been clearly demonstrated that under these conditions the decay rate is essentially only due to the long-range $1/R^2$ retardation term \cite{falkowski2026hitherto}.
This finding opens the door to a new class of environments and scenarios where ICD is relevant.

However, there are environments where the ICD of a single donor-acceptor pair is not energetically allowed, i.e., the excess energy of the donor is not sufficient to ionize the acceptor.
This raises the possibility of higher-order phenomena acting as alternative decay channels, such as collective ICD.
Recently, experimental studies revealed an intriguing phenomenon attributed to ICD in gas-phase monomers, where two or more monomer units collectively transfer their excitation energy to ionize an acceptor unit \cite{barik2022ambient,barik2023molecular,barik2026collective}.

The rate of collective ICD in multiply excited species was first formally derived by Cederbaum and Kuleff \cite{cederbaum2024collective}, building on the bare Coulomb interaction picture introduced earlier for such species \cite{kuleff2010ultrafast}.
They employed time-independent perturbation theory within non-relativistic quantum mechanics, considering the bare Coulomb interaction, which scales as $1/R^3$ in the coupling matrix element and yields a collective ICD rate proportional to $1/R^{12}$. Due to this short-range behavior of the rate, one anticipates that collective ICD is hard to explain in gases without including retardation.
Yet, how such retardation effects manifest themselves in collective ICD processes remains an open question. 

To answer this question, we report a novel theoretical framework based on time-independent perturbation theory within the interaction picture of non-relativistic quantum electrodynamics (QED). This approach incorporates retardation effects into collective ICD, extending the formulation of Ref.~\onlinecite{cederbaum2024collective} beyond the bare dipole-dipole Coulomb interaction. We shall also show that the theory developed enables one to take further relativistic effects into account.

\section{Theory}
\begin{figure}[!ht] 
	\centering
	\includegraphics[width=0.5\textwidth]{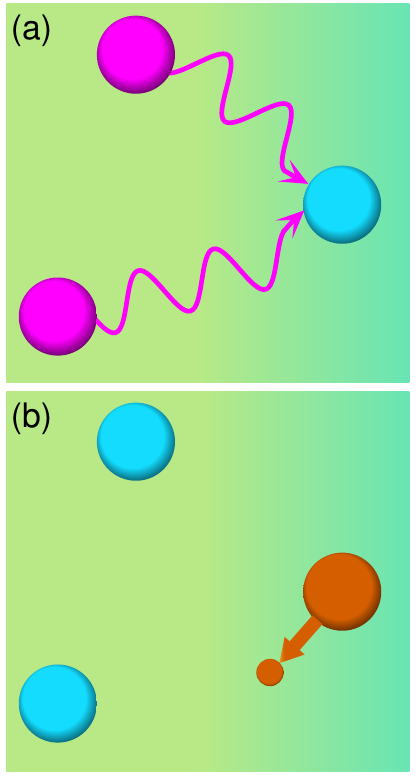}
	\caption{Illustration of collective ICD considering two donors and one acceptor. Panel (a) represents the process where both excited donors (magenta) emit virtual photons, which are absorbed by the acceptor. Panel (b) shows both donors in the ground state (cyan) and the ionized acceptor including the emitted ICD electron (dark orange).}
	\label{methods_fig_1}
\end{figure}

Consider a system composed of three particles, as illustrated in Figure \ref{methods_fig_1}. 
Initially, two excited donors $D_1^{*}$ and $D_2^{*}$ can transfer energy to an acceptor $A$ in the ground state.
Unlike the conventional ICD process, energy transfer from a single donor is insufficient to ionize the acceptor, but the combined excess energy of two donors is sufficient to do so.
In this case, we have
\begin{equation}
    D_1^* + D_2^* + A \rightarrow D_1 + D_2 + A^+
    \label{theory_eq_1}
\end{equation}
where $D_1$ and $D_2$ are the two de-excited donors and $A^{+}$ is the ionized acceptor, including the emitted ICD electron.
We are interested in evaluating the rate of the collective ICD from the energy transfer of two donors to the acceptor. In the following, we present our approach to describe the process in Eq. \eqref{theory_eq_1}, which consists of applying time-independent non-relativistic QED.

\subsection{Hamiltonian of the system}
The non-relativistic Hamiltonian of a system of electronic species (atoms or molecules) and an electromagnetic field in the Coulomb gauge is given by \cite{sakurai1967advanced,craig1998molecular}
\begin{equation}
    H = H_{mol} + H_{rad} + H_{int}
    \label{theory_eq_2}
\end{equation}
where $H_{mol} = \sum_{K} H_{K}$ is the Hamiltonian of all electronic species, $H_{rad}$ is the second quantized Hamiltonian of the radiation field, given by
\begin{equation}
     H_{rad} = \sum_{\vec{p},\lambda} \hbar \omega \left[a^{\dagger(\lambda)} (\vec{p}) a^{(\lambda)} (\vec{p}) + \frac{1}{2} \right]
    \label{theory_eq_3}
\end{equation}
with $a^{\dagger(\lambda)} (\vec{p})$ the creation and $a^{(\lambda)} (\vec{p})$ the annihilation operators of photons of polarization $\lambda$ and wave vector $\vec{p}$, $\omega = p c$ is the angular frequency, $\hbar$ the reduced Planck constant, and $c$ the speed of light.
The Hamiltonian $H_{int}$ describes the molecular interactions of the species and the field. By employing the Power-Zienau-Woolley transformation \cite{power1957radiative,woolley2020power}, it is possible to write the interaction Hamiltonian $H_{int}$ as \cite{craig1998molecular,salam2009molecular}
\begin{equation}
    H_{int} = - \frac{1}{\varepsilon_0} \sum_{K} \vec{d} (K) \cdot \vec{E^{\perp}} (\vec{R}_{K})
    \label{theory_eq_4}
\end{equation}
where $\vec{d} (K)$ is the dipole operator of the $K-\text{th}$ species and the transverse electric field $\vec{E^{\perp}}$ is given by
\begin{equation}
    \vec{E^{\perp}} (\vec{R}_{K}) = i \sum_{\vec{p},\lambda} \left(\frac{\hbar c p \varepsilon_0}{2 V}\right)^{1/2} \left[ \vec{\epsilon}^{(\lambda)} (\vec{p}) a^{(\lambda)} (\vec{p}) e^{+i \vec{p} \cdot \vec{R}_{K}} - \vec{\epsilon}^{(\lambda)*} (\vec{p}) a^{\dagger(\lambda)} (\vec{p}) e^{-i \vec{p} \cdot \vec{R}_{K}} \right]
    \label{theory_eq_5}
\end{equation}
with $\vec{R}_{K}$ the position of the species $K$, $\vec{\epsilon}^{(\lambda)}$ the polarization vector, $V$ is the volume of the quantized space and $\varepsilon_0$ the vacuum electric permittivity. 
The present approach is only valid for dipole-dipole interactions. For dark states and higher orders in the multipole expansion of the electric field, it is possible to go beyond by using the appropriate interaction Hamiltonian \cite{craig1998molecular,andrews2009resonance,salam2009molecular}. Here we restrict ourselves to using the electric dipole-dipole interaction Hamiltonian in Eq. \eqref{theory_eq_4}, which represents the leading term of intermolecular interactions of particles in bright states.

\subsection{Perturbation theory}
We are now in a position to discuss the perturbation theory approach to compute the collective ICD rate.
Both time-dependent and independent perturbation theory can be applied to calculate the transition rate since the result is independent of time \cite{salam2009molecular,sakurai2021modern}. 
We will proceed by considering the time-independent approach.

Following Eq. \eqref{theory_eq_2}, we adopt the standard perturbative notation $H = H_0 + W$, where we define the unperturbed system by $H_0 = H_{\text{mol}} + H_{\text{rad}}$, while the perturbation is represented by the interaction $W = H_{\text{int}}$.
The Schrödinger equation for the unperturbed system is given by
\begin{equation}
    H_0 | I \rangle = (H_{mol} + H_{rad}) | I \rangle  = E_I | I \rangle
    \label{methods_eq_4}
\end{equation}
where $|I\rangle$ denotes an unperturbed eigenstate, expressed as a tensor product of the atomic/molecular and radiation field states.
Formally, the unit operator of $H_0$ is given by
\begin{equation}
    \mathds{1}_{H_0} = \sum_{I} | I \rangle \langle I | = (\mathds{1}_{D_1} \otimes \mathds{1}_{D_2} \otimes \mathds{1}_{A}) \otimes \mathds{1}_{field}
    \label{methods_eq_5}
\end{equation}
The states of the participating species contain their ground states $| D_{1} \rangle$, $| D_{2} \rangle$, and $| A \rangle$ and all bound electronic excited states $| D_{1}^{m_1} \rangle$, $| D_{2}^{m_2} \rangle$ and $| A^{m_A} \rangle$ as well as all ionized states including the emitted electrons.  Of the latter, we restrict ourselves to the ionized state of the acceptor $| A^{+} \rangle$, which also includes the emitted collective ICD electron. The ionized states of the donors and those of the acceptor other than the state for which we compute the rate are not expected to contribute to the process. They can, if desired, be added \textit{a posteriori}. 
In addition, we consider for simplicity of presentation an acceptor which has no permanent dipole moment, like an atom or symmetric molecule.
The unit operators of the subspaces of each species given in Eq. \eqref{methods_eq_5} are written as
\begin{equation}
    \begin{split}
        \mathds{1}_{D_1} = & | D_{1} \rangle \langle D_{1} | + \sum_{m_1}^{\text{all excited}} | D_{1}^{(m_1)} \rangle \langle D_{1}^{(m_1)} | \\ 
        \mathds{1}_{D_2} = &  | D_{2} \rangle \langle D_{2} | + \sum_{m_2}^{\text{all excited}}| D_{2}^{(m_2)} \rangle \langle D_{2}^{(m_2)} |\\
        \mathds{1}_{A}   = &  | A \rangle \langle A | + \sum_{m_A}^{\text{all excited}}| A^{(m_A)} \rangle \langle A^{(m_A)} | + | A^{+} \rangle \langle A^{+} |\\
    \end{split}
    \label{methods_eq_6}
\end{equation}
The radiation field is defined by the virtual photons generated and annihilated through the emission and absorption processes of the electronic species. The corresponding field identity operator is given by
\begin{equation}
    \begin{split}
        &\mathds{1}_{field}  =  | 0\rangle \langle 0 | + 
        \sum_{\vec{p}_1,\lambda_1} | 1(\vec{p}_1,\lambda_1) \rangle \langle 1(\vec{p}_1,\lambda_1) | + \sum_{\vec{p}_2,\lambda_2} \sum_{\vec{p}_3,\lambda_3} | 1(\vec{p}_2,\lambda_2), 1(\vec{p}_3,\lambda_3) \rangle \langle 1(\vec{p}_2,\lambda_2), 1(\vec{p}_3,\lambda_3) |
    \end{split}
    \label{methods_eq_7}
\end{equation}
In Eq. \eqref{methods_eq_7}, the state space is truncated to include at most two photon modes (each containing one photon), accounting for the simultaneous presence of virtual photons from the donor species. More photons do not enter the calculation.

Having introduced the space of unperturbed states, we can define the initial state $| i \rangle$ and final state $| f \rangle$ (with energies $E_i$ and $E_f$, respectively) using the notation of Eqs. \eqref{methods_eq_5}, \eqref{methods_eq_6}, \eqref{methods_eq_7} as
\begin{equation}
    \begin{split}
    &| i \rangle = | D_{1}^{*}, D_{2}^{*}, A \rangle \otimes | 0  \rangle, \qquad  E_i = E_{D_1^*} + E_{D_2^*} + E_A, \\
    &| f \rangle = | D_{1}, D_{2}, A^{+} \rangle \otimes | 0  \rangle, \qquad  E_f = E_{D_1} + E_{D_2} + E_{A^{+}},
    \end{split}
    \label{theory_eq_6}
\end{equation}
where, due to energy conservation, $E_i = E_f$. This implies that the change of energy of the acceptor due to the process is $\Delta_A = E_{A^{+}} - E_{A} = \Delta_1 + \Delta_2$, where $\Delta_1 = E_{D_1^*} - E_{D_1}$ and $\Delta_2 =  E_{D_2^*} - E_{D_2}$ are the transferred excess energies by the donors $D_1$ and $D_2$, respectively.

The collective ICD rate $\Gamma$ can be computed using the generalized Fermi golden rule \cite{joachain1975quantum} 
\begin{equation}
    \Gamma = \frac{2 \pi}{\hbar} \rvert \langle f \lvert T \rvert i \rangle \rvert^2 
    \label{theory_eq_7}
\end{equation}
where the transition operator is \cite{sakurai2021modern} $T = W + W (E_i - H_0 + i0^{+})^{-1} T$. By using the Born series (recursion formula) \cite{cohen1997photons} we can rewrite $T$ using Eq. \eqref{methods_eq_4} and $W=H_{int}$ as
\begin{equation}
    \begin{split}
    T & = H_{int} +
    \sum_{I} \frac{ H_{int}| I \rangle \langle I | H_{int}}{(E_{i} - E_{I})} +
    \sum_{II} \sum_{I} \frac{ H_{int}| II \rangle \langle II |H_{int} | I \rangle \langle I |H_{int} }{(E_{i} - E_{II})(E_{i} - E_{I})} \\ & + \sum_{III} \sum_{II} \sum_{I} \frac{ H_{int}| III \rangle \langle III |H_{int} | II \rangle \langle II | H_{int} | I  \rangle \langle I | H_{int} }{(E_{i} - E_{III})(E_{i} - E_{II})(E_{i} - E_{I})} + \cdots  
    \end{split}
    \label{theory_eq_8}
\end{equation}
The intermediate states $|  I \rangle, |  II \rangle, |  III \rangle$ and the corresponding energies $E_{I}$, $E_{II}$, $E_{III}$ in the summations belong to the active space of the system defined by Eq. \eqref{methods_eq_5}.

\subsection{Transition amplitudes} 
The first non-zero term of the transition operator $T$ contributing to the collective energy transfer between two donors and one acceptor is of fourth order, as needed to describe the collective energy transfer between three particles mediated by a one-body potential ($H_{int}$). In the following, we shall concentrate on the evaluation of the respective matrix element of the transition operator in Eq. \eqref{theory_eq_8}:
\begin{equation}
    M_{fi} = \langle f \lvert T \rvert i \rangle = \sum_{III} \sum_{II} \sum_{I} \frac{\langle f | H_{int} | III \rangle \langle III  |  H_{int}  |  II \rangle \langle II  |  H_{int}  |  I \rangle \langle I  |  H_{int}  |  i \rangle}{(E_{i} - E_{III})(E_{i} - E_{II})(E_{i} - E_{I})}
    \label{theory_eq_9}
\end{equation}
The procedure to obtain the $M_{fi}$ is laborious. The details are given in Appendix \ref{appendix_A}. 
The total transition amplitude is calculated to consist of the following seven terms
\begin{equation}
        M_{fi} = M_{fi}^{(1)} + M_{fi}^{(2)} + M_{fi}^{(3)} + M_{fi}^{(4)} + M_{fi}^{(5)} + M_{fi}^{(6)} + M_{fi}^{(7)}
    \label{terms_eq_1}
\end{equation}
where
\begin{equation}
    \begin{split}
        M_{fi}^{(1)} = - \sum_{r} \sum_{s} \sum_{t} \sum_{u}
        d_{r}^{D_1 D_1^{*}} V_{rs} (k_1, \vec{R}_{D_{1}A}) \alpha_{st}^{A} (k_1, k_2) V_{tu} (k_2, \vec{R}_{D_{2}A}) d_{u}^{D_2 D_2^{*}} 
    \end{split}
    \label{terms_eq_2}
\end{equation}
\begin{equation}
        M_{fi}^{(2)} = - \frac{1}{\Delta_1 + \Delta_2} \sum_{r} \sum_{s} \sum_{t} \sum_{u}
        d_{r}^{\delta D_1} V_{rs} (k_1 + k_2, \vec{R}_{D_{1}A}) d_{s}^{A^{+} A} d_{t}^{D_1 D_1^{*}} V_{tu} (k_2, \vec{R}_{D_{1}D_{2}}) d_{u}^{D_2 D_2^{*}} 
    \label{terms_eq_3}
\end{equation}
\begin{equation}
        M_{fi}^{(3)} = \frac{1}{\Delta_2} \sum_{r} \sum_{s} \sum_{t} \sum_{u} d_{r}^{D_1 D_1^{*}} V_{rs} (k_1 + k_2, \vec{R}_{D_{1}A}) d_{s}^{A^{+} A} d_{t}^{\delta D_1} V_{tu} (k_2, \vec{R}_{D_{1}D_{2}}) d_{u}^{D_2 D_2^{*}} 
    \label{terms_eq_4}
\end{equation}
\begin{equation}
        M_{fi}^{(4)} = - \frac{1}{\Delta_1 + \Delta_2} \sum_{r} \sum_{s} \sum_{t} \sum_{u}
        d_{r}^{\delta D_2} V_{rs} (k_1 + k_2, \vec{R}_{D_{2}A}) d_{s}^{A^{+} A} d_{t}^{D_1 D_1^{*}} V_{tu} (k_1, \vec{R}_{D_{1}D_{2}}) d_{u}^{D_2 D_2^{*}} 
    \label{terms_eq_5}
\end{equation}
\begin{equation}
        M_{fi}^{(5)} = \frac{1}{\Delta_1} \sum_{r} \sum_{s} \sum_{t} \sum_{u} d_{r }^{D_2 D_2^{*}} V_{rs} (k_1 + k_2, \vec{R}_{D_{2}A}) d_{s}^{A^{+} A} d_{t}^{D_1 D_1^{*}} V_{tu} (k_1, \vec{R}_{D_{1}D_{2}}) d_{u}^{\delta D_2} 
    \label{terms_eq_6}
\end{equation}
\begin{equation}
        M_{fi}^{(6)} = - \sum_{r} \sum_{s} \sum_{t} \sum_{u}
         d_{r}^{A^{+} A} V_{rs} (k_1 + k_2, \vec{R}_{D_{1}A}) \alpha_{st}^{D_{1}} (- k_2, k_1 + k_2) V_{tu} (k_2, \vec{R}_{D_{1}D_{2}}) d_{u}^{D_2 D_2^{*}} 
    \label{terms_eq_7}
\end{equation}
\begin{equation}
        M_{fi}^{(7)} = - \sum_{r} \sum_{s} \sum_{t} \sum_{u}
         d_{r}^{A^{+} A} V_{rs} (k_1 + k_2, \vec{R}_{D_{2}A}) \alpha_{st}^{D_{2}} (- k_1, k_1 + k_2) V_{tu} (k_1, \vec{R}_{D_{1}D_{2}}) d_{u}^{D_1 D_1^{*}} 
    \label{terms_eq_8}
\end{equation}
In Eqs. \eqref{terms_eq_2}-\eqref{terms_eq_8}, the vector connecting the center of mass of the donor $D_1$ with that of $D_2$ is $\vec{R}_{D_{1}D_{2}}$ and analogously the vector connecting one donor and the acceptor is $\vec{R}_{D_{i}A}$ ($i=1,2$).
Here, the wave numbers $k_1$ and $k_2$ are defined via the donor excess energies $\Delta_1 = \hbar c k_1$ and $\Delta_2 = \hbar c k_2$ for $D_1$ and $D_2$, respectively.
A matrix element of the dipole moment operator is given by $\vec{d}^{S_1 S_2} = \langle S_1 \lvert \vec{d} \rvert S_2 \rangle$, where $|S_1\rangle$ and $|S_2 \rangle$ correspond to any state of the species defined in Eq. \eqref{methods_eq_6}. 
Interestingly, the difference between the permanent dipole moment in the excited state $|D_i^*\rangle$ and that in the ground state $|D_i\rangle$ of a donor also appears and reads $\vec{d}^{\delta D_i} = \vec{d}^{D_i^* D_i^*} - \vec{d}^{D_i D_i}$ ($i = 1, 2$).
The indices $r, s, t, u$ refer to Cartesian components of the quantities appearing in the equations. In addition to the matrix elements of the dipole moment operator, two new types of quantities appear:
\begin{equation}
    \alpha_{st}^{A} (k_1, k_2) = \sum_{n_A} \left (\frac{d_{s}^{A A^{(n_A)}} d_{t}^{A^{(n_A)} A^+} }{k_A^{(n_A)} - k_1} + \frac{d_{s}^{A^+ A^{(n_A)}} d_{t} ^{A^{(n_A)} A}}{k_A^{(n_A)} - k_2} \right), \quad \Delta_A^{(n_A)} = \hbar c k_A^{(n_A)} = E_A^{(n_A)} - E_A
    \label{terms_eq_9}
\end{equation}
\begin{equation}
    \alpha_{st}^{D_1} (- k_2, k_1 + k_2) = \sum_{n_1 \neq *} \left( \frac{ d_{s}^{D_1^{(n_1)} D_1} d_{t}^{D_1^* D_1^{(n_1)}}}{k_1^{(n_1)} + k_2} + \frac{d_{s}^{D_1^* D_1^{(n_1)}} d_{t}^{D_1^{(n_1)} D_1}}{k_1^{(n_1)} - k_1 - k_2} \right), \quad \Delta_{1}^{(n_1)} = \hbar c k_1^{(n_1)} = E_{D_1^{(n_1)}} - E_{D_1}
    \label{terms_eq_10}
\end{equation}
\begin{equation}
    \alpha_{st}^{D_2} (- k_1, k_1 + k_2) = \sum_{n_2 \neq *} \left( \frac{ d_{s}^{D_2^{(n_2)} D_2} d_{t}^{D_2^* D_2^{(n_2)}}}{k_2^{(n_2)} + k_1} + \frac{d_{s}^{D_2^* D_2^{(n_2)}} d_{t}^{D_2^{(n_2)} D_2}}{k_2^{(n_2)} - k_1 - k_2} \right), \quad \Delta_{2}^{(n_2)} = \hbar c k_2^{(n_2)} = E_{D_2^{(n_2)}} - E_{D_2^{}}
    \label{terms_eq_11}
\end{equation}
and
\begin{equation}
    V_{rs} (k, \vec{R}) = \frac{e^{ikR}}{4 \pi \varepsilon_0 R^3} 
    \left[
    (1 - ikR)(\delta_{rs} - 3 \hat{R}_r \hat{R}_s)
    -(kR)^2(\delta_{rs} - \hat{R}_r \hat{R}_s)  \right]
    \label{terms_eq_12}
\end{equation}
These quantities warrant further discussion. The term $\alpha_{st}^{K}$ (with $K \in \{D_1, D_2, A\}$) represents the transition polarizability tensor. For the acceptor ($K=A$), it describes a transition from the ground state $|A\rangle$ to the ionized state $|A^{+}\rangle$, mediated by all accessible bound states $|A^{(n_A)}\rangle$. For the donors ($K=D_1, D_2$), it describes a transition from the excited state $|D_i^*\rangle$ to the ground state $|D_i\rangle$, involving all possible bound states $|D_i^{(n_i)}\rangle$ as intermediate states (excluding the initially excited state $|D_i^*\rangle$ itself). Furthermore, $V_{rs}(k, \vec{R})$ denotes the retarded dipole-dipole interaction kernel mediated by a virtual photon of energy $\Delta = \hbar c k$. Finally, we mention that $\hat{R}$ is the unit vector in the direction of $\vec{R}$, i.e., $\hat{R}=\vec{R}/R$ and $R = |\vec{R}|$.

\subsection{Compact diagrammatic representation}
\label{diagrammatic_representation}

\begin{figure}[!ht] 
	\centering
	\includegraphics[width=1.0\textwidth]{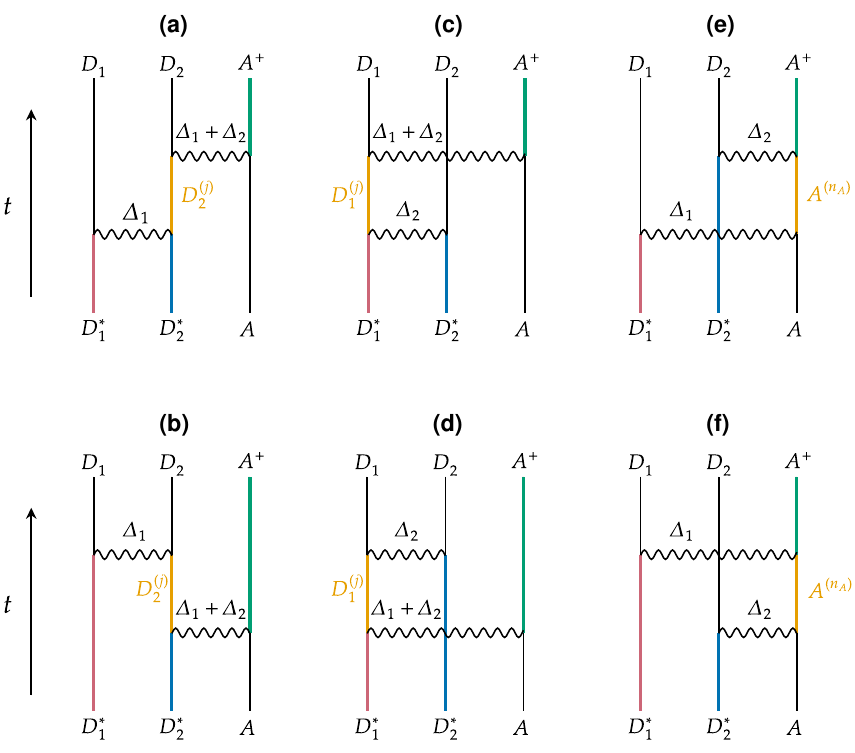}
	\caption{Diagrammatic representation of the collective ICD considering two donors ($D_1$ and $D_2$) and one acceptor ($A$). The initial state is described by the initially excited donors and the acceptor is in its ground state, i.e., $ | D_{1}^{*}, D_{2}^{*}, A \rangle$. In the final state, the donors are in the ground state and the acceptor is ionized, namely, $| D_{1}, D_{2}, A^{+} \rangle$. Diagrams (a) and (b) rely on the mediation of energy transfer by the donor $D_2$, which stores the energy $\Delta_1 = \hbar c k_1$ from $D_1$ in an intermediate state $|D_2^{(j)} \rangle$ and transfers the combined energy of $\Delta_1 + \Delta_2$ ($\Delta_2 = \hbar c k_2$) to the acceptor. Notice that $|D_2^{(j)}\rangle$ can be $|D_2\rangle$, $|D_2^{(n_2)}\rangle$ or $|D_2^*\rangle$. Similarly, diagrams (c) and (d) correspond to the same mediation process by the donor $D_1$. Diagrams (e) and (f) represent the process where both donors transfer their excess energy ($D_1$ transfers $\hbar c k_1$ and $D_2$ transfers $\hbar c k_2$) directly to the acceptor, which can be excited to an intermediate state $|A^{(n_A)} \rangle$. 
    }
	\label{methods_fig_2}
\end{figure}

The calculation of the transition amplitudes shown in Eqs. \eqref{terms_eq_2} -- \eqref{terms_eq_8} require involved procedures, independently of whether the derivation is performed by the insertion of intermediate states as described in the preceding subsection or via the evaluation of fourth-order Feynman diagrams.
In the latter case, a total of 7 diagrams with 24 time-orderings each resulting in 168 connected diagrams arise, which we have also evaluated, giving the same final result. Is it possible to use the same strategy employed in many-body theory of fermionic and bosonic systems where the interaction of two particles appears in the diagrams \cite{shavitt2009many,schirmer2018many} and thus reduce the number of diagrams substantially and simplify their evaluation? The answer is yes, and results in an elegant approach for collective energy transfer involving three bodies, as depicted in Figure \ref{methods_fig_2}.

First, it is important to note that since we are dealing with a two-body interaction potential, the diagrams in Figure \ref{methods_fig_2} are of second order instead of the conventional QED diagrams resulting from the one-body interaction $H_{int}$ which are of fourth order. The diagrammatic representation in Figure \ref{methods_fig_2} provides us with time-independent amplitudes. Regarding the time-ordering of the physical processes, as is done in QED Feynman diagrams, diagrams (b), (d), and (f) are the time-reversed ordering of diagrams (a), (c), and (e), respectively. Below, we define rules to construct and evaluate the diagrams, in addition to the fact that they must be connected diagrams. Time is represented on the vertical axis, flowing from bottom to top, and the position of the species is held fixed.

It is necessary to define and represent the initial and final states of the process, as well as the interactions between the species involved in the process. The initial state consists of the excited donors and the acceptor in the ground state: $|D_1^*, D_2^*, A \rangle$. In the diagrams of Figure \ref{methods_fig_2}, we highlight excited states with thicker, colored lines (red for donor $D_1$ and blue for donor $D_2$). The two-body interactions are represented by a wavy line characterized by the transferred energy directly above it. Each diagram starts at the bottom with the indices characterizing the initial state, followed by an interaction wavy line, which is then followed by an intermediate state of one species over which the summation runs and is indicated by a yellow line in the figure.
For the donors, the states can be the ground state or any of the excited states. For the acceptor, the intermediate states can be all possible bound excited states. After the second interaction, the system reaches the final state $| D_1, D_2, A^{+} \rangle$, in which the donors are in the ground state and the acceptor is ionized (green line), as represented at the top of each diagram.
But what is the interaction potential between the two species? 
Looking carefully at the tensor $V_{rs} (k, \vec{R})$ given in Eq. \eqref{terms_eq_12}, we can extract an interaction potential $\hat{W} (k, \vec{R})$ by summing over the Cartesian coordinates together with the dipole operators. Then, we can write 
\begin{equation}
    \begin{split}
    \hat{W} (k, \vec{R}) & =  \sum_{r} \sum_{s} d_{r} V_{rs} (k, \vec{R}) d_{s}' \\
    & = \frac{e^{ikR}}{4 \pi \varepsilon_0 R^3} 
    \left\{
    (1 - ikR)\left[\vec{d} \cdot \vec{d'} - 3 (\vec{d} \cdot \hat{R})(\vec{d'} \cdot \hat{R})\right]
    -(kR)^2\left[\vec{d} \cdot \vec{d'} - (\vec{d} \cdot \hat{R})(\vec{d'} \cdot \hat{R})\right]  \right\}        
    \end{split}
    \label{diag_eq_1}
\end{equation}
where $\Delta = \hbar c k$ is the energy transfer, $\vec{R}$
the vector connecting the centers of mass of the two species at distance $R = |\vec{R}|$, $\hat{R} = \vec{R}/R$ is the respective unit vector, and the operators $\vec{d}$ and $\vec{d'}$ are the dipole operators of the species. $\hat{W}$ is exactly the interaction between two species that one obtains from a relativistic treatment of the electron-electron interaction \cite{cederbaum2025relativistic} for small transferred energies.

Having introduced the rules to draw the diagrams in Figure 2, we are now in a position to extract the information from them.
The information is a transition amplitude of second-order perturbation theory, which takes on the appearance
\begin{equation}
    \text{\textbf{transition amplitude}} = \frac{(\text{\textbf{first interaction}}) \times (\text{\textbf{second interaction}})}{(\text{\textbf{initial energy}}) - (\text{\textbf{intermediate state energy}})}
    \label{eq0}
\end{equation}
Consider the diagram in Figure \ref{methods_fig_2}(a) as an example. As we already know from Eq. \eqref{theory_eq_6}, the initial state is $| D_{1}^{*}, D_{2}^{*}, A \rangle$ and the \textbf{initial energy} $= E_{D_1^*} + E_{D_2^{*}} + E_{A}$.
The first interaction (first wavy line from the bottom of the diagram) is between the donors in their initial state $| D_1^{*}, D_2^{*} \rangle$, where $D_1^{*}$ transfers its excess energy $\Delta_1 = \hbar c k_1$ to the donor $D_2^{*}$, leaving these species in $D_1$ and $D_2^{(j)}$, respectively. The intermediate state $| D_2^{(j)} \rangle$ can be the ground state or any bound excited state. We can write \textbf{first interaction} $ = \langle D_1^{*} , D_2^{*}| \hat{W} (k_1, \vec{R}_{D_1 D_2}) | D_1 , D_2^{(j)} \rangle$.
By drawing a hypothetical horizontal line between the first and second interaction wavy lines, we intermediately identify intermediate state of the system as $| D_1, D_2^{(j)}, A \rangle$ with $\text{\textbf{intermediate state energy}} = E_{D_1} + E_{D_2^{(j)}} + E_{A}$.
As seen in Figure \ref{methods_fig_2}(a), the second wavy line describes the interaction between the states $| D_2^{(j)}, A \rangle$ and $| D_2, A^{+} \rangle$, where the donor now transfers the total excess energy $\Delta_1 + \Delta_2$ ($\Delta_2 = \hbar c k_2$) to the acceptor $A$ ionizing it. We can thus identify the respective interaction as $\text{\textbf{second interaction}} = \langle D_2^{(j)} , A | \hat{W} (k_1 + k_2, \vec{R}_{D_2 A}) | D_2 , A^{+}\rangle$.
Having collected all the ingredients of the diagram in Figure \ref{methods_fig_2}(a), the resulting \textbf{transition amplitude} is given by
\begin{equation}
    M_{if}^{(a)} \left\{D_2^{(j)}\right\} = \frac{\langle D_1^{*} , D_2^{*}| \hat{W} (k_1, \vec{R}_{D_1 D_2}) | D_1 , D_2^{(j)} \rangle \langle D_2^{(j)} , A | \hat{W} (k_1 + k_2, \vec{R}_{D_2 A}) | D_2 , A \rangle}{(E_{D_1^*} + E_{D_2^{*}} + E_{A}) - (E_{D_1} + E_{D_2^{(j)}} + E_{A})}
    \label{diag_eq_2}
\end{equation}
In the following, we evaluate this contribution by considering the various possible intermediate states $| D_2^{(j)} \rangle$. To obtain the final contribution of the diagram, one has, of course, to sum over all possible intermediate states. Choosing $D_2^{(j)}  =  D_2$ and employing Eq. \eqref{diag_eq_1}, one obtains the explicit expression
\begin{equation}
    M_{fi}^{(a)} \left\{D_2 \right\} = \sum_{r} \sum_{s} \sum_{t} \sum_{u}
    \frac{d_{r}^{D_1 D_1^{*}} V_{rs} (k_1, \vec{R}_{D_{1}D_{2}}) d_{s}^{D_2 D_2^{*}} d_{t}^{D_2 D_2} V_{tu} (k_1 + k_2, \vec{R}_{D_{2}A}) d_{u}^{A A^{+}}}{\Delta_1 + \Delta_2}
    \label{diag_eq_3}
\end{equation}
By applying the same strategy as described above for the time-reversed diagram of Figure \ref{methods_fig_2}(b), we obtain
\begin{equation}
    M_{if}^{(b)} \left\{D_2^{(j)}\right\} = \frac{\langle D_2^{*} , A| \hat{W} (k_1 + k_2, \vec{R}_{D_2 A}) | D_2^{(j)} , A^{+} \rangle \langle D_1^{*}, D_2^{(j)} | \hat{W} (k_1, \vec{R}_{D_1 D_2}) | D_1, D_2 \rangle}{(E_{D_1^*} + E_{D_2^{*}} + E_{A}) - (E_{D_1^{*}} + E_{D_2^{(j)}} + E_{A^{+}})}
    \label{diag_eq_4}
\end{equation}
Choosing the case $D_2^{(j)} = D_2^{*}$, we can rewrite Eq. \eqref{diag_eq_4} in the same way as done for Eq. \eqref{diag_eq_3} to obtain
\begin{equation}
    M_{fi}^{(b)} \left\{D_2^{*}\right\} = \sum_{r} \sum_{s} \sum_{t} \sum_{u}
    \frac{d_{r}^{D_2^{*} D_2^{*}} V_{rs} (k_1 + k_2, \vec{R}_{D_{2}A}) d_{s}^{A A^{+}} d_{t}^{D_1 D_1^{*}} V_{tu} (k_1, \vec{R}_{D_{1}D_{2}}) d_{u}^{D_2 D_2^{*}} }{-\Delta_1 - \Delta_2}
    \label{diag_eq_5}
\end{equation}
By summing the contributions of Eqs. \eqref{diag_eq_3} and \eqref{diag_eq_5}, and comparing with Eq. \eqref{terms_eq_5} one can immediately see that
\begin{equation}
    M_{if}^{(a)}  \left\{D_2 \right\} + M_{if}^{(b)}  \left\{D_2^{*}\right\} = M_{if}^{(4)}
    \label{diag_eq_6}
\end{equation}
The amplitudes extracted from the other choices of $D_2^{(j)}$ as well as from the remaining diagrams in Figure \ref{methods_fig_2} are presented in Appendix \ref{appendix_B} together with their correspondence to Eqs. \eqref{terms_eq_2}--\eqref{terms_eq_8}. Clearly, the diagrams in Figure \ref{methods_fig_2} reproduce all the terms we have cumbersomely derived using QED. Altogether, we see that drawing and evaluating the diagrams in Figure \ref{methods_fig_2} results in a massive simplification of the original problem, reducing the 168 one-body QED interaction diagrams to just 6 two-body interaction diagrams.

It is important to interpret the physical meaning of the new diagrams. As discussed above, the diagrams in Figures \ref{methods_fig_2}(a) and \ref{methods_fig_2}(b) describe the process in which donor $D_2$ mediates the collective energy transfer to the acceptor. This mediation can occur either through the energy $\Delta_1$ transferred by $D_1^{*}$ and stored in $D_2^{*}$ via permanent dipoles (for $D_2^{(j)} = D_2 $ or $D_2^{*}$), or through the transition polarizability $\alpha_{st}^{D_2}$ (with $D_2^{(j)} = D_2^{(n_2)}$), where $n_2$ stands for all other excited states of this donor. The combined energy $\Delta_1 + \Delta_2$ is then transferred to the acceptor $A$, ionizing it.
Similarly, Figures \ref{methods_fig_2}(c) and \ref{methods_fig_2}(d) also represent a mediation process, but now mediated by donor $D_1$.

The diagrams in Figures \ref{methods_fig_2}(e) and \ref{methods_fig_2}(f) show that the two donors successively transfer their excess energy to the acceptor. The first transfer is accompanied by an excitation of the acceptor and the second transfer promotes the acceptor from the bound state $| A^{(n_A)} \rangle$ to the continuum state $| A^{+} \rangle$. From this interpretation, the transition polarizability tensor $\alpha_{st}^{A}$ acts as an energy storage for the acceptor, until the second excess energy is transferred and promotes the acceptor to the continuum.

\subsection{Averaging over orientations}
\label{averaging_over_orientations}
Having derived the amplitudes and understood their physical meaning, we can now proceed to obtain the collective ICD rate.

If the molecules involved are fixed in space, this rate follows straightforwardly from the golden rule in Eq. \eqref{theory_eq_7} and the matrix elements of the transition operator listed in Eqs. \eqref{theory_eq_9}--\eqref{terms_eq_8}. However, for most applications one needs the rate averaged over the orientation of the molecules. In the following, the averaging is carried out employing a concise technique. The averaged rates are much more amenable to calculations and to interpretations as they are rather compact and, as we shall see, related to measurable quantities, see also Ref. \onlinecite{cederbaum2024collective}. 

With the techniques we shall use, all terms of the collective ICD rate can be averaged over the molecular orientations. However, as we consider the terms of the transition amplitudes which depend on the permanent dipole moments to be more relevant than those containing transition polarizabilities (see also Ref. \onlinecite{cederbaum2024collective}) and for convenience of presentation, we proceed by considering only the respective amplitudes, i.e., the amplitudes in Eqs. \eqref{terms_eq_3}--\eqref{terms_eq_6}.

By using Eq. \eqref{theory_eq_7}, the total rate is given by
\begin{equation}
    \begin{split}
        \Gamma = \text{ } & \frac{2 \pi}{\hbar} \Biggl\{\lvert M_{fi}^{(2)} \rvert^2 + \lvert M_{fi}^{(3)} \rvert^2 + \lvert M_{fi}^{(4)} \rvert^2 + \lvert M_{fi}^{(5)} \rvert^2 + 2 \Re\left[M_{fi}^{(2)} (M_{fi}^{(3)})^{*}\right] + 2 \Re\left[M_{fi}^{(2)} (M_{fi}^{(4)})^{*}\right] \\
        & + 2 \Re\left[M_{fi}^{(2)} (M_{fi}^{(5)})^{*}\right] + 2 \Re\left[M_{fi}^{(3)} (M_{fi}^{(4)})^{*}\right] + 2 \Re\left[M_{fi}^{(3)} (M_{fi}^{(5)})^{*}\right] + 2 \Re\left[M_{fi}^{(4)} (M_{fi}^{(5)})^{*}\right] \Biggr\}        
    \end{split}
    \label{avg_eq_1}
\end{equation}
By defining $\Gamma^{(v,w)} = (2 \pi/\hbar) M_{fi}^{(v)} (M_{fi}^{(w)})^{*}$, we have
\begin{equation}
        \Gamma = \frac{2 \pi}{\hbar} \sum_{v = 2}^{5} \sum_{w = 2}^{5} M_{fi}^{(v)} (M_{fi}^{(w)})^{*} = \sum_{v = 2}^{5} \sum_{w = 2}^{5} \Gamma^{(v,w)}
    \label{avg_eq_2}
\end{equation}
For electronic species in isotropic environments such as liquids, gases, and solutions, averaging over orientations (or rotational averaging) is an inevitable procedure for obtaining isotropic invariant observables. Only after averaging over orientations we can relate the rate to measurable quantities. Denoting $\langle \Gamma \rangle$ as the averaged-over-orientations rate, from Eq. \eqref{avg_eq_2}, we have
\begin{equation}
        \langle \Gamma \rangle = \frac{2 \pi}{\hbar} \sum_{v = 2}^{5} \sum_{w = 2}^{5} \langle M_{fi}^{(v)} (M_{fi}^{(w)})^{*} \rangle = \sum_{v = 2}^{5} \sum_{w = 2}^{5} \langle \Gamma^{(v,w)} \rangle
    \label{avg_eq_3}
\end{equation}
We now need to obtain the individual rates $\langle \Gamma^{(v,w)} \rangle$.
By inspecting Eqs. \eqref{terms_eq_2} -- \eqref{terms_eq_8}, we can write a general equation for $\langle \Gamma^{(v,w)} \rangle$ as \cite{craig1998molecular,bonvicini2023threedim}
\begin{equation}
    \langle \Gamma^{(v,w)} \rangle = \sum_{i_1 ... i_8}  A_{i_1 ... i_8}^{(v,w)} \langle P_{i_1 ... i_8}^{(v,w)} \rangle
    \label{avg_eq_5}
\end{equation}
where $A_{i_1 ... i_8}$ represents the product of interaction potential tensors $V_{rs} (k, \vec{R})$ in fixed positions $\vec{R}$ (relative distance between species) and $P_{i_1 ... i_8}$ the product of molecular properties (for example, permanent or transition dipole moments), both written with the Cartesian components $i_1 ... i_8$ in the laboratory frame (LF). The number of Cartesian components is eight since we have four Cartesian coordinate indices from $M_{fi}^{(v)}$ plus another four from $(M_{fi}^{(w)})^{*}$.
However, molecular properties are usually computed in the body frame (BF), and from Eq. \eqref{avg_eq_5}, it is necessary to represent them in the LF. Then, by considering the Cartesian components in the BF as $\lambda_1 ... \lambda_8$, we have
\begin{equation}
    \langle \Gamma^{(v,w)} \rangle = \sum_{i_1 ... i_8}  A_{i_1 ... i_8}^{(v,w)} \sum_{\lambda_1 ... \lambda_8} \langle l_{i_{1} \lambda_1} ... l_{i_{8} \lambda_8} \rangle P_{\lambda_1 ... \lambda_8}^{(v,w)}
    \label{avg_eq_6}
\end{equation}
where the quantity $l_{i_{1} \lambda_1} ... l_{i_{8} \lambda_8}$ is the so-called direction cosines product of angles between the BF and LF for Cartesian tensors of rank $8$. Each term $l_{i_{p} \lambda_p}$ is the $(i_{p},\lambda_p)$ element of the $3 \times 3$ Euler angle matrix, and it is parameterized through the Euler angles $(\phi, \theta, \psi)$. The isotropic average $\langle l_{i_{1} \lambda_1 ... i_{n} \lambda_n} \rangle$ can be represented by \cite{bonvicini2023threedim}
\begin{equation}
    I^{(n)} = \langle l_{i_{1} \lambda_1} ... l_{i_{n} \lambda_n} \rangle = \frac{1}{8 \pi^2} \int_{\phi=0}^{2 \pi} \int_{\theta=0}^{\pi} \int_{\psi=0}^{2 \pi} l_{i_{1} \lambda_1} ... l_{i_{n} \lambda_n} \sin{\theta} d \phi d\theta d\psi
    \label{avg_eq_7}
\end{equation}
For $n = 2, 3, 4$, is possible to show that \cite{craig1998molecular}
\begin{equation}
    I^{(2)} = \frac{1}{3} \delta_{i_1 i_2} \delta_{\lambda_1 \lambda_2}
    \label{avg_eq_8}
\end{equation}
\begin{equation}
    I^{(3)} = \frac{1}{6} \varepsilon_{i_1 i_2 i_3} \varepsilon_{\lambda_1 \lambda_2 \lambda_3}
    \label{avg_eq_9}
\end{equation}
\begin{equation}
    I^{(4)} = \frac{1}{30} 
    \begin{pmatrix}
        \delta_{i_1 i_2} \delta_{i_3 i_4} \\
        \delta_{i_1 i_3} \delta_{i_2 i_4} \\
        \delta_{i_1 i_4} \delta_{i_2 i_3} \\
    \end{pmatrix}
    ^{\mathbf{T}}
    \begin{pmatrix}
         4 & -1 & -1 \\
        -1 &  4 & -1 \\
        -1 & -1 &  4 \\
    \end{pmatrix}
    \begin{pmatrix}
        \delta_{\lambda_1 \lambda_2} \delta_{\lambda_3 \lambda_4} \\
        \delta_{\lambda_1 \lambda_3} \delta_{\lambda_2 \lambda_4} \\
        \delta_{\lambda_1 \lambda_4} \delta_{\lambda_2 \lambda_3} \\
    \end{pmatrix}
    \label{avg_eq_10}
\end{equation}

Now we have all the tools to obtain $\langle \Gamma \rangle$. Let us consider, as an example, the rate associated with the transition amplitude $M_{if}^{(4)}$ in Eq. \eqref{terms_eq_5} using Eq. \eqref{avg_eq_5}
\begin{equation}
    \begin{split}
         \langle \Gamma^{(4,4)} \rangle = \frac{2 \pi }{\hbar(\Delta_1 + \Delta_2)^2}
         & \sum_{r,s,t,u}  \sum_{r',s',t',u'}
        \langle d_{r}^{\delta D_2} d_{r'}^{\delta D_2} d_{u}^{D_2 D_2^{*}} d_{u'}^{D_2 D_2^{*}} \rangle \langle d_{s}^{A^{+} A} d_{s'}^{A^{+} A} \rangle \langle d_{t}^{D_1 D_1^{*}} d_{t'}^{D_1 D_1^{*}} \rangle \\
        \times &V_{rs} (k_1 + k_2, \vec{R}_{D_{2}A}) V_{r's'}^{*} (k_1 + k_2, \vec{R}_{D_{2}A}) V_{tu} (k_1, \vec{R}_{D_{1}D_{2}}) V_{t'u'}^{*} (k_1, \vec{R}_{D_{1}D_{2}}) 
    \end{split}
    \label{avg_eq_11}
\end{equation}
where $r,s,t,u$ and $r',s',t',u'$ are indices regarding Cartesian components in the LF. In Eq. \eqref{avg_eq_11}, we grouped the molecular properties that belong to the same species. By using Eqs. \eqref{avg_eq_6}, \eqref{avg_eq_8}, \eqref{avg_eq_9} and \eqref{avg_eq_10}, it is possible to obtain
\begin{equation}
        \langle d_{s}^{A^{+} A} d_{s'}^{A^{+} A}  \rangle =  \frac{1}{3} \lvert \vec{d}^{A^{+} A} \rvert^2 \delta_{s s'}, \qquad  \langle d_{t}^{D_1 D_1^{*}} d_{t'}^{D_1 D_1^{*}}  \rangle =  \frac{1}{3} \lvert \vec{d}^{D_1 D_1^{*}} \rvert^2 \delta_{t t'}
    \label{avg_eq_12}
\end{equation}
\begin{equation}
    \begin{split}
        \langle d_{r}^{\delta D_2} d_{u}^{D_2 D_2^{*}} d_{r'}^{\delta D_2} d_{u'}^{D_2 D_2^{*}} \rangle = \frac{1}{30} \Biggl\{ & \left[4 \left( \vec{d}^{\delta D_2} \cdot \vec{d}^{D_2 D_2^{*}}\right)^2 - 2 \lvert \vec{d}^{\delta D_2} \rvert^2 \lvert \vec{d}^{D_2 D_2^{*}}  \rvert^2 \right]  \delta_{r u}  \delta_{r' u'}  \\
        - & \left[\left( \vec{d}^{\delta D_2} \cdot \vec{d}^{D_2 D_2^{*}}\right)^2 - 3 \lvert \vec{d}^{\delta D_2} \rvert^2 \lvert \vec{d}^{D_2 D_2^{*}}  \rvert^2 \right] \left(\delta_{r u'} \delta_{u r'} + \delta_{r r'} \delta_{u u'} \right) \Biggr\}
    \end{split}
    \label{avg_eq_13}
\end{equation}
Inserting Eqs. \eqref{avg_eq_12} and \eqref{avg_eq_13} in Eq. \eqref{avg_eq_11} we obtain
\begin{equation}
    \begin{split}
         \langle \Gamma^{(4,4)} \rangle = \frac{2 \pi }{270 \hbar (\Delta_1 + \Delta_2)^2}
         & \sum_{r,s,t,u}  \sum_{r',s',t',u'}
        \lvert \vec{d}^{D_1 D_1^{*}} \rvert^2 \lvert \vec{d}^{A^{+} A} \rvert^2 \delta_{s s'} \delta_{t t'} \\
         \times \Biggl\{ &  \left[4 \left( \vec{d}^{\delta D_2} \cdot \vec{d}^{D_2 D_2^{*}}\right)^2 - 2 \lvert \vec{d}^{\delta D_2} \rvert^2 \lvert \vec{d}^{D_2 D_2^{*}}  \rvert^2 \right]  \delta_{r u}  \delta_{r' u'} \\
        - & \left[\left( \vec{d}^{\delta D_2} \cdot \vec{d}^{D_2 D_2^{*}}\right)^2 - 3 \lvert \vec{d}^{\delta D_2} \rvert^2 \lvert \vec{d}^{D_2 D_2^{*}}  \rvert^2 \right] \left(\delta_{r u'} \delta_{u r'} + \delta_{r r'} \delta_{u u'} \right) \Biggr\}\\
        \times &V_{rs} (k_1 + k_2, \vec{R}_{D_{2}A}) V_{r's'}^{*} (k_1 + k_2, \vec{R}_{D_{2}A}) V_{tu} (k_1, \vec{R}_{D_{1}D_{2}}) V_{t'u'}^{*} (k_1, \vec{R}_{D_{1}D_{2}}) 
    \end{split}
    \label{avg_eq_14}
\end{equation}
For simplicity, considering the far-zone region ($V_{rs} (k, \vec{R}) \propto  k^2(\delta_{rs} - \hat{R}_r \hat{R}_s) / R $, see Eq. \eqref{terms_eq_12}), after some algebra it is possible to show that Eq. \eqref{avg_eq_14} reduces to
\begin{equation}
    \begin{split}
        \langle \Gamma^{(4,4)} \rangle = & \frac{2 \pi }{270 \hbar (\Delta_1 + \Delta_2)^2} \frac{k_1^4 (k_1 + k_2)^4}{(4 \pi \varepsilon_0)^4 R_{D_2 A}^2 R_{D_1 D_2}^2} \lvert \vec{d}^{D_1 D_1^{*}} \rvert^2 \lvert \vec{d}^{A^{+} A} \rvert^2 \\
         \times \Biggl\{ &\left( \vec{d}^{\delta D_2} \cdot \vec{d}^{D_2 D_2^{*}}\right)^2 \left[- 1 + 3 \left( \hat{R}_{D_{2}A} \cdot \hat{R}_{D_{1}D_{2}} \right)^2 \right] + \lvert \vec{d}^{\delta D_2} \rvert^2 \lvert \vec{d}^{D_2 D_2^{*}}  \rvert^2 \left[ 13 + \left( \hat{R}_{D_{2}A} \cdot \hat{R}_{D_{1}D_{2}}\right)^2 \right] \Biggl\}
    \end{split}
    \label{avg_eq_15}
\end{equation}

The procedure to obtain the other terms in Eq. \eqref{avg_eq_2} is similar to that done for Eq. \eqref{avg_eq_15}. All the terms in Eq. \eqref{avg_eq_2} are listed in Appendix \ref{appendix_C}, where more details about the averaging techniques can also be found. The total rate reads
\begin{equation}
    \begin{split}
        & \langle \Gamma \rangle = \frac{2 \pi }{270 \hbar (4 \pi \varepsilon_0)^4(\hbar c)^8 } \frac{(\Delta_1 + \Delta_2)^4}{ R_{D_1 D_2}^2} \lvert \vec{d}^{D_1 D_1^{*}}  \rvert^2 \lvert \vec{d}^{D_2 D_2^{*}} \rvert^2 \lvert \vec{d}^{A^{+} A} \rvert^2 \\
        \times \Biggl\{ & \left[\frac{1}{(\Delta_1 + \Delta_2)^2}+ \frac{1}{(\Delta_2)^2} \right]\frac{\Delta_2^4}{R_{D_1 A}^2} \lvert \vec{d}^{\delta D_1} \rvert^2 \Biggl\{ \left( \hat{d}^{\delta D_1} \cdot \hat{d}^{D_1 D_1^{*}}\right)^2 \left[- 1 + 3 \left( \hat{R}_{D_{1}A} \cdot \hat{R}_{D_{1}D_{2}} \right)^2 \right] +  \left[ 13 + \left( \hat{R}_{D_{1}A} \cdot \hat{R}_{D_{1}D_{2}}\right)^2 \right] \Biggl\} \\
        + & \left[\frac{1}{(\Delta_1 + \Delta_2)^2} + \frac{1}{(\Delta_1)^2} \right]\frac{\Delta_1^4}{R_{D_2 A}^2} \lvert \vec{d}^{\delta D_2} \rvert^2 \Biggl\{ \left( \hat{d}^{\delta D_2} \cdot \hat{d}^{D_2 D_2^{*}}\right)^2 \left[- 1 + 3 \left( \hat{R}_{D_{2}A} \cdot \hat{R}_{D_{1}D_{2}} \right)^2 \right] +  \left[ 13 + \left( \hat{R}_{D_{2}A} \cdot \hat{R}_{D_{1}D_{2}}\right)^2 \right] \Biggl\} \\
        - &  \frac{2\Delta_2^4}{\Delta_2(\Delta_1 + \Delta_2)R_{D_1 A}^2} \lvert \vec{d}^{\delta D_1} \rvert^2 
         \Biggl\{ \left[- 1 + 3 \left( \hat{R}_{D_{1}A} \cdot \hat{R}_{D_{1}D_{2}} \right)^2 \right] +  \left( \hat{d}^{\delta D_1} \cdot \hat{d}^{D_1 D_1^{*}}\right)^2 \left[ 13 + \left( \hat{R}_{D_{1}A} \cdot \hat{R}_{D_{1}D_{2}}\right)^2 \right] \Biggl\} \\
        - & \frac{2 \Delta_1^4}{\Delta_1 (\Delta_1 + \Delta_2)R_{D_2 A}^2} \lvert \vec{d}^{\delta D_2} \rvert^2 
         \Biggl\{ \left[- 1 + 3 \left( \hat{R}_{D_{2}A} \cdot \hat{R}_{D_{1}D_{2}} \right)^2 \right] +  \left( \hat{d}^{\delta D_2} \cdot \hat{d}^{D_2 D_2^{*}}\right)^2 \left[ 13 + \left( \hat{R}_{D_{2}A} \cdot \hat{R}_{D_{1}D_{2}}\right)^2 \right] \Biggl\} \Biggl\}
    \end{split}
    \label{avg_eq_16}
\end{equation}
Notice that some mixed terms vanish
\begin{equation}
    \begin{split}
        \langle \Gamma^{(2,4)} \rangle = \langle \Gamma^{(4,2)} \rangle = \langle \Gamma^{(2,5)} \rangle = \langle \Gamma^{(5,2)} \rangle = \langle \Gamma^{(3,5)} \rangle = \langle \Gamma^{(5,3)} \rangle = 0
    \end{split}
    \label{avg_eq_17}
\end{equation}
which is due to a product of three tensors that leads to a determinant with two equal rows/columns in the Levi Civita symbol in Eq. \eqref{avg_eq_9}, for example $\langle d_{r}^{\delta D_2} d_{s}^{\delta D_2} d_{t}^{D_2 D_2^{*}} \rangle = 0$.

\begin{figure}[!h] 
	\centering
	\includegraphics[width=0.5\textwidth]{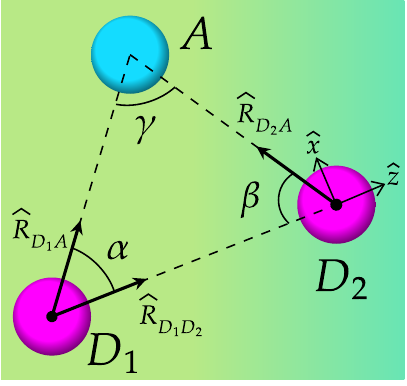}
	\caption{Illustration of the geometric configuration in the collective ICD of two donors (magenta) and one acceptor (cyan).}
	\label{methods_fig_3}
\end{figure}
In Eq. \eqref{avg_eq_17}, scalar products of unit vectors connecting the species appear. These can be expressed in terms of geometric quantities, as illustrated in Figure \ref{methods_fig_3}. The species reside in the plane $xz$ in Cartesian coordinates, and the unit vectors between species are
\begin{equation}
    \begin{split}
        &\hat{R}_{D_1 D_2} = \hat{z} \\
        &\hat{R}_{D_1 A} = \cos (\alpha) \hat{x} + \sin (\alpha) \hat{z} \\
        &\hat{R}_{D_2 A} = \cos (\beta) \hat{x} - \sin (\beta) \hat{z}
    \end{split}
    \label{geom_eq_1}
\end{equation}
where the angle between $\hat{R}_{D_1 A}$ and $\hat{R}_{D_2 A}$ is $\gamma$, and for a triangle we have $\alpha + \beta + \gamma = \pi$. 

\subsection{Donors of the same electronic species}

Many environments possess donors of the same electronic species.
In this case, the energies and the magnitudes of the change of the permanent dipole moments and of the transition dipole moments are equal, that is:
\begin{equation}
    \begin{split}
        & \Delta_1 = \Delta_2 \equiv \Delta \\
        & \lvert \vec{d}^{D_1 D_1^{*}} \rvert = \lvert \vec{d}^{D_2 D_2^{*}} \rvert \equiv \lvert \vec{d}^{D D^{*}} \rvert \\
        & \lvert \vec{d}^{\delta D_1} \rvert = \lvert \vec{d}^{\delta D_2} \rvert \equiv \lvert\vec{d}^{\delta D} \rvert \\
        & \hat{d}^{\delta D_1} \cdot \hat{d}^{D_1 D_1^{*}} = \hat{d}^{\delta D_2} \cdot \hat{d}^{D_2 D_2^{*}} \equiv \hat{d}^{\delta D} \cdot \hat{d}^{D D^{*}} = \cos (\eta) 
    \end{split}
    \label{geom_eq_2}
\end{equation}
By using Eqs. \eqref{geom_eq_1} and \eqref{geom_eq_2} in the total rate in Eq. \eqref{avg_eq_16}, we obtain the compact and informative expression
\begin{equation}
    \begin{split}
        \langle \Gamma \rangle = & \frac{2 \pi }{270 \hbar (4 \pi \varepsilon_0)^4(\hbar c)^8 } \frac{2^4 \Delta^6}{4 R_{D_1 D_2}^2} \lvert \vec{d}^{D D^{*}}  \rvert^4 \lvert \vec{d}^{A^{+} A} \rvert^2 \lvert \vec{d}^{\delta D} \rvert^2 \left(\frac{g_1}{R_{D_1 A}^2}  +  
        \frac{g_2}{R_{D_2 A}^2} \right)
    \end{split}
    \label{geom_eq_3}
\end{equation}
where the auxiliary functions $g_1$ and $g_2$ read
\begin{equation}
    \begin{split}
        & g_1 = \cos^2 (\eta) \left[- 57 + 11 \sin^2 (\alpha) \right] +  \left[ 69 - 7 \sin^2 (\alpha) \right] \\
        & g_2 = \cos^2 (\eta) \left[- 57 + 11 \sin^2 (\beta) \right] +  \left[ 69 - 7 \sin^2 (\beta) \right]
    \end{split}
    \label{geom_eq_4}
\end{equation}

There are cases, for example, in planar molecules, where the direction between the permanent dipole moment $\vec{d}^{\delta D}$ and the transition dipole moment $\vec{d}^{D D^{*}}$ is parallel or perpendicular.
For the parallel case, this means that $(\hat{d}^{\delta D} \cdot \hat{d}^{D D^{*}})^2 = \cos^2 (\eta)$ is equal to 1, while, when the directions are perpendicular, $\cos^2 (\eta) = 0$. 
Considering this in the auxiliary functions in Eq. \eqref{geom_eq_4}, the total rate in Eq. \eqref{geom_eq_3}, for these two cases, takes on the following explicit forms
\begin{equation}
    \begin{split}
        & \langle \Gamma \rangle_{\parallel} = \frac{2 \pi }{270 \hbar (4 \pi \varepsilon_0)^4(\hbar c)^8 } \frac{2^4 \Delta^6}{4 R_{D_1 D_2}^2} \lvert \vec{d}^{D D^{*}}  \rvert^4 \lvert \vec{d}^{A^{+} A} \rvert^2 \lvert \vec{d}^{\delta D} \rvert^2 \left[\frac{16 - 4 \cos^2 (\alpha)}{R_{D_1 A}^2}  +  
        \frac{16 - 4 \cos^2 (\beta)}{R_{D_2 A}^2} \right] \\
        & \langle \Gamma \rangle_{\perp} = \frac{2 \pi }{270 \hbar (4 \pi \varepsilon_0)^4(\hbar c)^8 } \frac{2^4 \Delta^6}{4 R_{D_1 D_2}^2} \lvert \vec{d}^{D D^{*}}  \rvert^4 \lvert \vec{d}^{A^{+} A} \rvert^2 \lvert \vec{d}^{\delta D} \rvert^2 \left[\frac{62 + 7 \cos^2 (\alpha)}{R_{D_1 A}^2}  +  
        \frac{62 + 7 \cos^2 (\beta)}{R_{D_2 A}^2} \right]
    \end{split}
    \label{geom_eq_5}
\end{equation}
It is clear that $\langle \Gamma \rangle_{\perp} > \langle \Gamma \rangle_{\parallel}$, which means that the collective ICD process is more efficient for donors with initially excited states where the vectors $\vec{d}^{\delta D}$ and $\vec{d}^{D D^{*}}$ are perpendicular to each other.

\subsection{Expressing the ICD rate by measurable quantities}

We are now in a position to express the collective ICD rate in terms of measurable quantities.
The photoionization cross section of the acceptor $\sigma_{pi}$ is expressed as \cite{sobel1972introduction,cederbaum2024collective}
\begin{equation}
    \sigma_{pi}^{A} = \frac{4 \pi^2 (2\Delta)}{3c\hbar} \lvert \vec{d}^{A^{+} A} \rvert^2
    \label{measurable_eq_1}
\end{equation}
As also done by Cederbaum and Kuleff \cite{cederbaum2024collective} for the Coulomb interaction ($1/R^3$ term, see Eq. \eqref{terms_eq_12}), we can obtain the spontaneous emission rate (the Einstein coefficient) of a photon with energy $2\Delta$ collectively emitted by the donors.
It is possible to show that \cite{falkowski2026einsteincoeff}
\begin{equation}
    A_{2} = \frac{(2 \Delta)^3}{3 \pi \varepsilon_0 \hbar^4 c^3} \lvert \mathcal{T} \rvert^2
    \label{measurable_eq_2}
\end{equation}
where, by using the quantities defined in Eq. \eqref{geom_eq_2}, we have
\begin{equation}
    \lvert \mathcal{T} \rvert^2 = \frac{2 \Delta^2 \lvert \vec{d}^{\delta D} \rvert^2 \lvert \vec{d}^{D D^{*}}  \rvert^4}{9(4\pi \varepsilon_0)^2 (\hbar c)^4 R_{D_1 D_2}^2}  g_3
    \label{measurable_eq_3}
\end{equation}
with the auxiliary function $g_3 = [(5 / 4) - \cos^2(\eta)]$. 
For the parallel case ($\vec{d}^{\delta D} \parallel \vec{d}^{D D^{*}}$) $g_3 = 1/4$ and for the perpendicular case ($\vec{d}^{\delta D} \perp \vec{d}^{D D^{*}}$) $g_3 = 5/4$.
By inserting Eqs. \eqref{measurable_eq_1} -- \eqref{measurable_eq_3} in Eq. \eqref{geom_eq_5}, we obtain the final far-zone result
\begin{equation}
    \begin{split}
        & \langle \Gamma \rangle_{\parallel} = \frac{1}{(4 \pi \varepsilon_0)} \frac{3}{40} \left[\frac{16 - 4 \cos^2 (\alpha)}{R_{D_1 A}^2}  +  
        \frac{16 - 4 \cos^2 (\beta)}{R_{D_2 A}^2} \right] \frac{A_2^{\parallel} \sigma_{pi}^A}{4 \pi} \\
        & \langle \Gamma \rangle_{\perp} = \frac{1}{(4 \pi \varepsilon_0)} \frac{3}{200} \left[\frac{62 + 7 \cos^2 (\alpha)}{R_{D_1 A}^2}  +  
        \frac{62 + 7 \cos^2 (\beta)}{R_{D_2 A}^2} \right] \frac{A_2^{\perp} \sigma_{pi}^A}{4 \pi} 
    \end{split}
    \label{measurable_eq_4}
\end{equation}
The rates in Eq. \eqref{measurable_eq_4} show that the collective ICD rate does not depend explicitly on the transferred energies, as in the case of the ICD rate with one donor and one acceptor in the far-zone \cite{hemmerich2018influence,cederbaum2025relativistic}. It depends only on the measurable quantities $A_2$ and $\sigma_{pi}^A$, the distances between the species and their relative geometric orientation (the angles $\alpha$ and $\beta$ in Figure \ref{methods_fig_3}).

\section{Conclusions}

In standard ICD, the excess energy of a single donor is utilized to ionize a neighboring acceptor. If the excess energy is insufficiently large to perform ICD, but the collective transfer of the excess energy of two donors suffices, we encounter collective ICD. In the case that the donor and acceptor are rather close to each other, like in clusters and liquids, it has been shown that the accurate ICD rate is orders of magnitude larger than that predicted by the asymptotic expansion in powers of the inverse distance between the donor and acceptor \cite{averbukh2004mechanism}. This finding underlines the true power of ICD and we expect a similar behavior also for collective ICD. 

More recently, it became clear that ICD and collective ICD can also take place if the distances between the participating species are large, like, for instance, in gases \cite{falkowski2026hitherto,barik2022ambient}. It became evident that this is only possible if many donors and acceptors are present and, in particular, retardation effects are included \cite{falkowski2026hitherto}. Retardation effects have been shown to be rather negligible in clusters and liquids.
For standard ICD, the operative expressions for the rates including retardation have been derived \cite{hemmerich2018influence,cederbaum2025relativistic} and applied for gases \cite{falkowski2026hitherto}. The aim of the present work has been to derive the working equations for collective ICD including retardation to be able to investigate the process also in gases and related environments. 

We presented an algebraic procedure using time-independent perturbation theory via a non-relativistic QED approach to evaluate the transition amplitudes and the rate for collective ICD, where two donors collectively transfer their excess energy to ionize an acceptor.  Particular attention is paid to the behavior of these quantities at large distances by going beyond the dipole-dipole Coulomb interaction between the species. The approach consists of explicitly writing the fourth-order transition amplitudes (the first non-vanishing order describing the process) and inserting completeness by the states of the unperturbed Hamiltonian, composed of the electronic-species part and the radiation field. We employed the Power-Zienau-Wooley transformed dipole interaction (see Eq. \eqref{theory_eq_4}) to represent the interaction Hamiltonian. A retarded interaction potential is identified in the calculations and represented by the conventional $1/R^3$ term, as well as the retardation terms $1/R^2$ and $1/R$, see Eq. \eqref{terms_eq_12}. The advantage of this procedure is its possible implementation in programming languages that support symbolic algebra, as well as in Wolfram Mathematica, expanding the applicability to systems with more species. 

The general transition amplitude obtained in Eqs. \eqref{terms_eq_2} -- \eqref{terms_eq_8} exhibits interesting physical aspects of the energy transfer from the donors to the acceptor. Among them, we highlight: (i) the difference between the permanent dipole moments of the donors in the excited and ground states plays an important role by acting as a bridge between the donors in their collective energy transfer, (ii) the transition polarizability of the acceptor acts as a storage of energy which can be utilized in the collective process, and (iii) the transition polarizability of the donors can also store excess energy. The role of the difference in permanent dipole moments was already reported by Cederbaum and Kuleff \cite{cederbaum2024collective} when describing collective ICD without retardation in the context of the dipole-dipole Coulomb interaction. 
Here, we find that this role also persists in the terms describing retardation.

An important result is the correspondence we find between the problem solved by the fourth-order QED perturbation theory and the diagrammatic representation using two-body interaction potentials as commonly used in the many-body perturbation theory. This correspondence massively simplifies the original problem (168 fourth-order connected diagrams in standard QED) by employing the diagrammatic representation in Figure \ref{methods_fig_2}, which gives rise to only 6 diagrams of second order. The final result of both approaches is shown to be identical, as exemplified in Subsection \ref{diagrammatic_representation}. This correspondence should enable one to attack collective processes of more excited donors where one expects many more terms in standard QED. It might also enable the use of relativistic potentials in collective ICD, as explored by Cederbaum and Hofierka \cite{cederbaum2025relativistic} by employing the Breit interaction in the transition amplitudes in the case of ICD of one donor and one acceptor.

In addition, by considering the common situation where the two donors are of the same electronic species, the expressions for the rates are strongly simplified. In particular, if the difference of permanent dipole moments discussed above is parallel or perpendicular to the transition dipole moment, we show that the resulting rate can be expressed explicitly by measurable quantities (the collective Einstein coefficient $A_2$  of the two donors and the photoionization cross section $\sigma_{pi}^A$ of the acceptor) and the geometric configuration of the species in space. We expect similar interpretations to hold also in the general case. 

One can envisage many situations where collective ICD can play a role in gases. In dilute gases, we refer to the recent experiments performed in Refs. \onlinecite{barik2022ambient,barik2023molecular,barik2026collective}, where molecules have been excited by a weak laser. In less dilute gases, like in the Earth's atmosphere and in particular in interstellar clouds near stars, low-energy excitation of gas constituents can lead to the appearance of unexpected ions of molecules with rather large ionization potentials.

\section*{Acknowledgments}
The authors thank Alexander I. Kuleff, Stefan Yoshi Buhmann and Jaroslav Hofierka for valuable discussions. Financial support by the Deutsche Forschungsgemeinschaft (DFG) (Grant No. CE 10/56-1) is gratefully acknowledged. A.G.F. acknowledges support from grants Nrs. 2024/17762-5 and 2025/07261-1, São Paulo Research Foundation (FAPESP).

\section*{Author Declarations}
\subsection*{Conflict of Interest}

The authors have no conflicts to disclose.

\subsection*{Author Contributions}
\noindent\textbf{Alan G. Falkowski:} Conceptualization (equal); Data curation (lead); Formal analysis (lead); Funding acquisition (equal); Investigation (lead); Methodology (lead); Visualization (lead); Writing – original draft (equal); Writing - review \& editing (equal); 
\textbf{Lorenz S. Cederbaum:} Conceptualization (equal); Supervision (lead); Funding acquisition (equal);  Writing – original draft (equal); Writing – review \& editing (equal).

\section*{Data Availability Statment}
The data that support the findings of this study are available from the corresponding author upon reasonable request.

\appendix
\section{Derivation of the QED fourth-order transition amplitudes}
\label{appendix_A}
Deriving Eqs. \eqref{terms_eq_2}--\eqref{terms_eq_8} is lengthy. For the identity operator in Eq. \eqref{methods_eq_5}, which spans the unperturbed states of the system, we first consider two arbitrary states, $|S_1 \rangle = |\text{mol}_1 \rangle \otimes |\text{rad}_1 \rangle$ and $|S_2 \rangle = |\text{mol}_2 \rangle \otimes |\text{rad}_2 \rangle$. Here, $|\text{mol}_1 \rangle$ and $|\text{mol}_2 \rangle$ denote the electronic states of the species, while $|\text{rad}_1 \rangle$ and $|\text{rad}_2 \rangle$ correspond to the radiation field.
Then, by using Eq. \eqref{methods_eq_4}, the matrix element of $H_{int}$ is 
\begin{equation}
    \langle S_1 | H_{int} | S_2 \rangle = -\frac{1}{\varepsilon_0} \sum_{K}  \langle \text{mol}_1 | \vec{d} (K) | \text{mol}_2 \rangle \cdot \langle \text{rad}_1 | \vec{E^{\perp}} (\vec{R}_{K}) | \text{rad}_2 \rangle
    \label{appendix_A_eq_1}
\end{equation}
Therefore, we can separate the evaluation of the matrix elements of $H_{int}$ into contributions from the electronic species and from the radiation field, which will help to evaluate $M_{fi}$ more effectively.

Listing all possible states of the donors and acceptor, we have
\begin{equation}
    \begin{aligned}
    | m_1 \rangle &= | D_1 , D_2 , A \rangle, & | m_{10} \rangle &= | D_1^{*} , D_2^{*} , A \rangle, & | m_{19} \rangle &= | D_1^{(n_1)} , D_2^{(n_2)} , A \rangle \\
    | m_2 \rangle &= | D_1 , D_2 , A^{(n_A)} \rangle, & | m_{11} \rangle &= | D_1^{*} , D_2^{*} , A^{(n_A)} \rangle, & | m_{20} \rangle &= | D_1^{(n_1)} , D_2^{(n_2)} , A^{(n_A)} \rangle \\
    | m_3 \rangle &= | D_1 , D_2 , A^{+} \rangle, & | m_{12} \rangle &= | D_1^{*} , D_2^{*} , A^{+} \rangle, & | m_{21} \rangle &= | D_1^{(n_1)} , D_2^{(n_2)} , A^{+} \rangle \\
    | m_4 \rangle &= | D_1^{*} , D_2 , A \rangle, & | m_{13} \rangle &= | D_1^{(n_1)} , D_2 , A \rangle, & | m_{22} \rangle &= | D_1^{(n_1)} , D_2^{*} , A \rangle \\
    | m_5 \rangle &= | D_1^{*} , D_2 , A^{(n_A)} \rangle, & | m_{14} \rangle &= | D_1^{(n_1)} , D_2 , A^{(n_A)} \rangle, & | m_{23} \rangle &= | D_1^{(n_1)} , D_2^{*} , A^{(n_A)} \rangle \\
    | m_6 \rangle &= | D_1^{*} , D_2 , A^{+} \rangle, & | m_{15} \rangle &= | D_1^{(n_1)} , D_2 , A^{+} \rangle, & | m_{24} \rangle &= | D_1^{(n_1)} , D_2^{*} , A^{+} \rangle \\
    | m_7 \rangle &= | D_1 , D_2^{*} , A \rangle, & | m_{16} \rangle &= | D_1 , D_2^{(n_2)} , A \rangle, & | m_{25} \rangle &= | D_1^{*} , D_2^{(n_2)} , A \rangle \\
    | m_8 \rangle &= | D_1 , D_2^{*} , A^{(n_A)} \rangle, & | m_{17} \rangle &= | D_1 , D_2^{(n_2)} , A^{(n_A)} \rangle, & | m_{26} \rangle &= | D_1^{*} , D_2^{(n_2)} , A^{(n_A)} \rangle \\
    | m_9 \rangle &= | D_1 , D_2^{*} , A^{+} \rangle, & | m_{18} \rangle &= | D_1 , D_2^{(n_2)} , A^{+} \rangle, & | m_{27} \rangle &= | D_1^{*} , D_2^{(n_2)} , A^{+} \rangle
    \end{aligned}
    \label{appendix_A_eq_2}
\end{equation}
where $n_1, n_2 \neq * $.
A generic matrix element of the dipole operator $\sum_{K} \vec{d} (K)$ is given by
\begin{equation}
    \begin{split}
        \langle D_1^{(r)}, D_2^{(s)}, A^{(t)} | \sum_{K} \vec{d} (K) | D_1^{(u)}, D_2^{(v)}, A^{(w)} \rangle & =  \langle D_1^{(r)}| \vec{d} (D_1) | D_1^{(u)} \rangle \delta_{sv} \delta_{tw} \\
        & + \langle D_2^{(s)}| \vec{d} (D_2) | D_2^{(v)} \rangle \delta_{ru} \delta_{tw} \\
        & + \langle A^{(t)} | \vec{d} (A) | A^{(w)} \rangle \delta_{ru} \delta_{sv} \\
        & =  \vec{d}^{D_1^{(r)} D_1^{(u)}} \delta_{sv} \delta_{tw} + \vec{d}^{D_2^{(s)} D_2^{(v)}} \delta_{ru} \delta_{tw} + \vec{d}^{A^{(t)} A^{(w)}} \delta_{ru} \delta_{sv}
    \end{split}
    \label{appendix_A_eq_3} 
\end{equation}
where we have made use of the orthonormality of the unperturbed states of all species.

For the radiation field, the unit operator defined in Eq. \eqref{methods_eq_7} has three states
\begin{equation}
    | 0 \rangle, \quad  |  1(\vec{p}_1, \lambda_1) \rangle, \quad | 1(\vec{p}_2,\lambda_2), 1(\vec{p}_3,\lambda_3) \rangle
    \label{appendix_A_eq_5}
\end{equation}
Using the rules for creation and annihilation operators for bosons \cite{sakurai2021modern} and from Eqs. \eqref{theory_eq_5} and \eqref{methods_eq_7}, the non-zero matrix elements of $\vec{E^{\perp}} (\vec{R}_{K})$ are
\begin{equation}
    \begin{aligned}
        \langle 1(\vec{p}_1, \lambda_1) | \vec{E^{\perp}} (\vec{R}_{K}) | 0 \rangle = - i \left(\frac{\hbar c p_1 \varepsilon_0}{2 V}\right)^{1/2} \vec{\epsilon}^{(\lambda)*} (\vec{p}_1) e^{-i \vec{p}_1 \cdot \vec{R}_{K} } 
    \end{aligned}
    \label{appendix_A_eq_6}    
\end{equation}
\begin{equation}
    \begin{aligned}
        \langle 1(\vec{p}_1,\lambda_1), 1(\vec{p}_2,\lambda_2) | \vec{E^{\perp}} (\vec{R}_{K}) | 1(\vec{p}_3,\lambda_3) \rangle= - i \left(\frac{\hbar c p_2 \varepsilon_0}{2 V}\right)^{1/2} \vec{\epsilon}^{(\lambda_2)*} (\vec{p}_2) e^{-i \vec{p}_2 \cdot \vec{R}_{K}}  \delta_{\vec{p}_1 \vec{p}_3} \delta_{\lambda_1 \lambda_3}  \\ - i \left(\frac{\hbar c p_1 \varepsilon_0}{2 V}\right)^{1/2} \vec{\epsilon}^{(\lambda_1)*} (\vec{p}_1) e^{-i \vec{p}_1 \cdot \vec{R}_{K}} \delta_{\vec{p}_2 \vec{p}_3} \delta_{\lambda_2 \lambda_3}
    \end{aligned}
    \label{appendix_A_eq_7}    
\end{equation}
and $\langle 0 | \vec{E^{\perp}} (\vec{R}_{K}) | 1(\vec{p}_1, \lambda_1)  \rangle = \langle 1(\vec{p}_1, \lambda_1) | \vec{E^{\perp}} (\vec{R}_{K}) | 0 \rangle^{\dagger}$ and $\langle 1(\vec{p}_3,\lambda_3) | \vec{E^{\perp}} (\vec{R}_{K}) | 1(\vec{p}_1,\lambda_1), 1(\vec{p}_2,\lambda_2)  \rangle=\langle 1(\vec{p}_1,\lambda_1), 1(\vec{p}_2,\lambda_2) | \vec{E^{\perp}} (\vec{R}_{K}) | 1(\vec{p}_3,\lambda_3) \rangle^{\dagger}$.

Notice that from Eqs. \eqref{theory_eq_6} and \eqref{appendix_A_eq_2}, we have $| i \rangle =| D_1^{*}, D_2^{*}, A \rangle \otimes | 0 \rangle  \equiv | m_{10} \rangle \otimes | 0 \rangle$ and $| f \rangle = | D_1, D_2, A^{+} \rangle \otimes | 0 \rangle  \equiv | m_3 \rangle \otimes | 0 \rangle$.
Therefore, by considering Eqs. \eqref{appendix_A_eq_1} -- \eqref{appendix_A_eq_5}, the transition amplitude in Eq. \eqref{theory_eq_9} can be written in three terms
\begin{equation}
    \begin{split}
    & M_{if}^{'} = (\varepsilon_0)^{-4} \sum_{m_{III}} \sum_{m_{II}} \sum_{m_{I}} \sum_{K_{1}} \sum_{K_{2}}  \sum_{\vec{p}_1,\lambda_1} \sum_{\vec{p}_2,\lambda_2} \sum_{r} \sum_{s} \sum_{t} \sum_{u} \\
    \times & \left(\frac{1}{E_i - E_{m_{III}} - \hbar \omega_1} \right) \left(\frac{1}{E_i - E_{m_{II}}}\right) \left(\frac{1}{E_i - E_{m_{I}} - \hbar \omega_2} \right) \\
    \times & \langle m_{3} | d_r (K_{1}) | m_{III} \rangle \langle m_{III} | d_s (K_{1}) | m_{II} \rangle \langle m_{II} | d_t (K_{2}) | m_{I} \rangle  \langle m_{I} | d_u (K_{2}) | m_{10} \rangle\\
    \times &   \langle 0 | E^{\perp}_r (\vec{R}_{K_1}) | 1(\vec{p}_1, \lambda_1) \rangle \langle 1(\vec{p}_1, \lambda_1) | E^{\perp}_s (\vec{R}_{K_1}) | 0 \rangle  \\
    \times & \langle 0 | E^{\perp}_t (\vec{R}_{K_2}) | 1(\vec{p}_2, \lambda_2) \rangle \langle 1(\vec{p}_2, \lambda_2) | E^{\perp}_u (\vec{R}_{K_2}) | 0 \rangle
    \end{split}
    \label{appendix_A_eq_8}
\end{equation}
\begin{equation}
    \begin{split}
    M_{if}^{''} = & (\varepsilon_0)^{-4} \sum_{m_{III}} \sum_{m_{II}} \sum_{m_{I}} \sum_{K_{1}} \sum_{K_{2}} \sum_{\vec{p}_1,\lambda_1} \sum_{\vec{p}_2,\lambda_2} \sum_{r} \sum_{s} \sum_{t} \sum_{u}\\
    \times & \left(\frac{1}{E_i - E_{m_{III}} - \hbar \omega_1} \right) \left(\frac{1}{E_i - E_{m_{II}} - \hbar \omega_1 - \hbar \omega_2}\right) \left(\frac{1}{E_i - E_{m_{I}} - \hbar \omega_2} \right) \\
    \times & \langle m_{3} | d_r (K_{1}) | m_{III} \rangle \langle m_{III} | d_s (K_{2}) | m_{II} \rangle \langle m_{II} | d_t (K_{1}) | m_{I} \rangle \langle m_{I} | d_u (K_{2}) | m_{10} \rangle  \\
    \times & \langle 0 | E^{\perp}_r (\vec{R}_{K_1}) | 1(\vec{p}_1, \lambda_1) \rangle \langle 1(\vec{p}_1, \lambda_1) | E^{\perp}_s (\vec{R}_{K_2}) | 1(\vec{p}_1, \lambda_1) , 1(\vec{p}_2, \lambda_2)\rangle \\ 
    \times & \langle 1(\vec{p}_1, \lambda_1) , 1(\vec{p}_2, \lambda_2)| E^{\perp}_t (\vec{R}_{K_1}) | 1(\vec{p}_2, \lambda_2) \rangle  \langle 1(\vec{p}_2, \lambda_2) | E^{\perp}_u (\vec{R}_{K_2}) | 0 \rangle
    \end{split}
    \label{appendix_A_eq_9}
\end{equation}
\begin{equation}
    \begin{split}
    M_{if}^{'''} = & (\varepsilon_0)^{-4} \sum_{m_{III}} \sum_{m_{II}} \sum_{m_{I}} \sum_{K_{1}} \sum_{K_{2}} \sum_{\vec{p}_1,\lambda_1} \sum_{\vec{p}_2,\lambda_2} \\
    \times & \left(\frac{1}{E_i - E_{m_{III}} - \hbar \omega_1} \right) \left(\frac{1}{E_i - E_{m_{II}} - \hbar \omega_1 - \hbar \omega_2}\right) \left(\frac{1}{E_i - E_{m_{I}} - \hbar \omega_1} \right) \\
    \times & \langle m_{3} | d_r (K_{1}) | m_{III} \rangle \langle m_{III} | d_s (K_{2}) | m_{II} \rangle \langle m_{II} | d_t (K_{2}) | m_{I} \rangle \langle m_{I} | d_u (K_{1}) | m_{10} \rangle  \\
    \times & \langle 0 | \vec{E^{\perp}} (\vec{R}_{K_1}) | 1(\vec{p}_1, \lambda_1) \rangle \langle 1(\vec{p}_1, \lambda_1) | \vec{E^{\perp}} (\vec{R}_{K_2}) | 1(\vec{p}_1, \lambda_1) , 1(\vec{p}_2, \lambda_2)\rangle \\ 
    \times & \langle 1(\vec{p}_1, \lambda_1) , 1(\vec{p}_2, \lambda_2)| \vec{E^{\perp}} (\vec{R}_{K_2}) | 1(\vec{p}_1, \lambda_1) \rangle  \langle 1(\vec{p}_1, \lambda_1) | \vec{E^{\perp}} (\vec{R}_{K_1}) | 0 \rangle
    \end{split}
    \label{appendix_A_eq_10}
\end{equation}
where $r, s, t, u$ are Cartesian components ($x, y, z$).

To simplify the evaluation of the transition amplitudes in Eqs. \eqref{appendix_A_eq_8}--\eqref{appendix_A_eq_10}, we first identify which intermediate states yield non-zero dipole matrix elements with the initial state $\vert{} D_1^{*}, D_2^{*}, A \rangle \equiv \vert{} m_{10} \rangle$ and the final state $\vert{} D_1, D_2, A^{+} \rangle \equiv \vert{} m_{3} \rangle$.
By using Eq. \eqref{appendix_A_eq_2} and Eq. \eqref{appendix_A_eq_3}, the non-zero terms elements regarding $| m_{10} \rangle$ (initial state) are
\begin{equation}
    \begin{split}
    \sum_{m_I} | m_I \rangle \langle m_I | \sum_{K} \vec{d} (K) | m_{10}\rangle =  \Bigl(&| m_4 \rangle \langle m_4 | + | m_7 \rangle \langle m_7 | + | m_{10} \rangle \langle m_{10} | + | m_{11} \rangle \langle m_{11} | \\ 
     + &| m_{12} \rangle \langle m_{12} | + | m_{22} \rangle \langle m_{22} | + | m_{25} \rangle\langle m_{25} |\Bigr) \sum_{K_1} \vec{d} (K_1) | m_{10} \rangle 
    \end{split}
    \label{appendix_A_eq_11}
\end{equation}
Similarly, for $| m_{3} \rangle$ (final state), we have
\begin{equation}
    \begin{split}
    \langle m_{3} | \sum_{K} \vec{d} (K) \sum_{m_{III}} | m_{III} \rangle \langle m_{III} | = & \langle m_{3} | \sum_{K} \vec{d} (K) \Bigl(| m_{1} \rangle \langle m_{1} | + | m_{2} \rangle \langle m_{2} | + | m_{3} \rangle \langle m_{3} | \\
    + &| m_{6} \rangle \langle m_{6} | + | m_{9} \rangle \langle m_{9} | + | m_{15} \rangle \langle m_{15} | + | m_{18} \rangle \langle m_{18} |\Bigr)
    \end{split}
    \label{appendix_A_eq_12}
\end{equation}

The transition amplitudes $M_{fi}'$, $M_{fi}''$, and $M_{fi}'''$ in Eqs. \eqref{appendix_A_eq_8}--\eqref{appendix_A_eq_10} are further evaluated by inserting Eqs. \eqref{appendix_A_eq_11} and \eqref{appendix_A_eq_12} and employing the explicit matrix elements from Eqs. \eqref{appendix_A_eq_3}, \eqref{appendix_A_eq_6}, and \eqref{appendix_A_eq_7}. The explicit expressions for these amplitudes are given in Appendix~\ref{appendix_D}.

\allowdisplaybreaks

Let us give an example of the calculation required. The transition amplitudes associated with the intermediate states $| A^{(n_A)} \rangle$ of the acceptor are
\begin{align}
& M_{fi}^{(1)} =  \sum_{r} \sum_{s} \sum_{t} \sum_{u} \sum_{n_A} \sum_{\vec{p}_1,\lambda_1} \sum_{\vec{p}_2,\lambda_2}\left(\frac{\hbar c p_1}{2 \epsilon_0 V}\right) \left(\frac{\hbar c p_2}{2 \epsilon_0 V}\right) \nonumber \\ \times \Biggl[ &
-\frac{e^{i \left(\vec{p}_2 \cdot \vec{R}_{D_{1} A}+\vec{p}_1 \cdot \vec{R}_{D_{2} A}\right)} \left( d_r^{D_{2} D_{2}^{*}} \epsilon_r^{(\lambda_1)}\right) \left( d_s^{D_{1} D_{1}^{*}} \epsilon_s^{(\lambda_2)}\right) \left( d_t^{A^{+} A^{(n_A)}} \epsilon_t^{(\lambda_2)}\right)  \left( d_u^{A^{(n_A)} A} \epsilon_u^{(\lambda_1)} \right)}{(-\Delta_{2}-\hbar \omega_1) (-\Delta_{A}^{(n_A)}-\hbar \omega_1) (\Delta_{1}+\Delta_{2}+\hbar \omega_1+\hbar \omega_2)}\nonumber \\ &-\frac{e^{i \left(\vec{p}_1 \cdot \vec{R}_{D_{1} A}+\vec{p}_2 \cdot \vec{R}_{D_{2} A}\right)} \left( d_r^{D_{1} D_{1}^{*}} \epsilon_r^{(\lambda_1)}\right) \left( d_s^{D_{2} D_{2}^{*}} \epsilon_s^{(\lambda_2)}\right) \left( d_t^{A^{+} A^{(n_A)}} \epsilon_t^{(\lambda_2)}\right)  \left( d_u^{A^{(n_A)} A} \epsilon_u^{(\lambda_1)} \right)}{(-\Delta_{1}-\hbar \omega_1) (-\Delta_{A}^{(n_A)}-\hbar \omega_1) (\Delta_{1}+\Delta_{2}+\hbar \omega_1+\hbar \omega_2)}\nonumber \\ &-\frac{e^{i \left(-\vec{p}_2 \cdot \vec{R}_{D_{1} A}+\vec{p}_1 \cdot \vec{R}_{D_{2} A}\right)} \left(d_r^{D_{2} D_{2}^{*}} \epsilon_r^{(\lambda_1)}\right) \left( d_s^{A^{+} A^{(n_A)}} \epsilon_s^{(\lambda_2)}\right) \left( d_t^{D_{1} D_{1}^{*}} \epsilon_t^{(\lambda_2)}\right)  \left( d_u^{A^{(n_A)} A} \epsilon_u^{(\lambda_1)} \right)}{(-\Delta_{2}-\hbar \omega_1) (-\Delta_{A}^{(n_A)}-\hbar \omega_1) (-\Delta_{1}+\Delta_{A}^{(n_A)}+\hbar \omega_1+\hbar \omega_2)}\nonumber \\ &-\frac{e^{i \left(\vec{p}_1 \cdot \vec{R}_{D_{1} A}-\vec{p}_2 \cdot \vec{R}_{D_{2} A}\right)} \left( d_r^{D_{1} D_{1}^{*}} \epsilon_r^{(\lambda_1)}\right) \left( d_s^{A^{+} A^{(n_A)}} \epsilon_s^{(\lambda_2)}\right) \left( d_t^{D_{2} D_{2}^{*}} \epsilon_t^{(\lambda_2)}\right)  \left( d_u^{A^{(n_A)} A} \epsilon_u^{(\lambda_1)} \right)}{(-\Delta_{1}-\hbar \omega_1) (-\Delta_{A}^{(n_A)}-\hbar \omega_1) (-\Delta_{2}+\Delta_{A}^{(n_A)}+\hbar \omega_1+\hbar \omega_2)}\nonumber \\ &-\frac{e^{-i \vec{p}_1 \cdot \vec{R}_{D_{1} A}+i \vec{p}_2 \cdot \vec{R}_{D_{2} A}} \left( d_r^{A^{+} A^{(n_A)}} \epsilon_r^{(\lambda_1)}\right) \left( d_s^{D_{2} D_{2}^{*}} \epsilon_s^{(\lambda_2)}\right) \left( d_t^{A^{(n_A)} A} \epsilon_t^{(\lambda_2)}\right)  \left( d_u^{D_{1} D_{1}^{*}}\epsilon_u^{(\lambda_1)} \right)}{(\Delta_{1}-\hbar \omega_1) (\Delta_{1}+\Delta_{2}-\Delta_{A}^{(n_A)}-\hbar \omega_1) (-\Delta_{1}+\Delta_{A}^{(n_A)}+\hbar \omega_1+\hbar \omega_2)}\nonumber \\ &-\frac{e^{-i \left(\vec{p}_1 \cdot \vec{R}_{D_{1} A}+\vec{p}_2 \cdot \vec{R}_{D_{2} A}\right)} \left( d_r^{A^{+} A^{(n_A)}} \epsilon_r^{(\lambda_1)}\right) \left( d_s^{A^{(n_A)} A} \epsilon_s^{(\lambda_2)}\right) \left( d_t^{D_{2} D_{2}^{*}} \epsilon_t^{(\lambda_2)}\right)  \left( d_u^{D_{1} D_{1}^{*}}\epsilon_u^{(\lambda_1)} \right)}{(\Delta_{1}-\hbar \omega_1) (\Delta_{1}+\Delta_{2}-\Delta_{A}^{(n_A)}-\hbar \omega_1) (-\Delta_{1}-\Delta_{2}+\hbar \omega_1+\hbar \omega_2)}\nonumber \\ &-\frac{e^{i \left(\vec{p}_2 \cdot \vec{R}_{D_{1} A}-\vec{p}_1 \cdot \vec{R}_{D_{2} A}\right)} \left( d_r^{A^{+} A^{(n_A)}} \epsilon_r^{(\lambda_1)}\right) \left( d_s^{D_{1} D_{1}^{*}}\epsilon_s^{(\lambda_2)}\right) \left( d_t^{A^{(n_A)} A} \epsilon_t^{(\lambda_2)}\right)  \left( d_u^{D_{2} D_{2}^{*}}\epsilon_u^{(\lambda_1)} \right)}{(\Delta_{2}-\hbar \omega_1) (\Delta_{1}+\Delta_{2}-\Delta_{A}^{(n_A)}-\hbar \omega_1) (-\Delta_{2}+\Delta_{A}^{(n_A)}+\hbar \omega_1+\hbar \omega_2)}\nonumber \\ &-\frac{e^{-i \left(\vec{p}_2 \cdot \vec{R}_{D_{1} A}+\vec{p}_1 \cdot \vec{R}_{D_{2} A}\right)} \left( d_r^{A^{+} A^{(n_A)}} \epsilon_r^{(\lambda_1)}\right) \left( d_s^{A^{(n_A)} A} \epsilon_s^{(\lambda_2)}\right) \left( d_t^{D_{1} D_{1}^{*}} \epsilon_t^{(\lambda_2)}\right)  \left( d_u^{D_{2} D_{2}^{*}}\epsilon_u^{(\lambda_1)} \right)}{(\Delta_{2}-\hbar \omega_1) (\Delta_{1}+\Delta_{2}-\Delta_{A}^{(n_A)}-\hbar \omega_1) (-\Delta_{1}-\Delta_{2}+\hbar \omega_1+\hbar \omega_2)}\nonumber \\ &-\frac{e^{i \left(\vec{p}_2 \cdot \vec{R}_{D_{1} A}+\vec{p}_1 \cdot \vec{R}_{D_{2} A}\right)} \left( d_r^{D_{2} D_{2}^{*}} \epsilon_r^{(\lambda_1)}\right) \left( d_s^{D_{1} D_{1}^{*}}\epsilon_s^{(\lambda_2)}\right) \left( d_t^{A^{+} A^{(n_A)}} \epsilon_t^{(\lambda_1)}\right)  \left( d_u^{A^{(n_A)} A}\epsilon_u^{(\lambda_2)} \right)}{(-\Delta_{2}-\hbar \omega_1) (-\Delta_{A}^{(n_A)}-\hbar \omega_2) (\Delta_{1}+\Delta_{2}+\hbar \omega_1+\hbar \omega_2)}\nonumber \\ &-\frac{e^{i \left(\vec{p}_1 \cdot \vec{R}_{D_{1} A}+\vec{p}_2 \cdot \vec{R}_{D_{2} A}\right)} \left(d_r^{D_{1} D_{1}^{*}} \epsilon_r^{(\lambda_1)}\right) \left( d_s^{D_{2} D_{2}^{*}} \epsilon_s^{(\lambda_2)}\right) \left( d_t^{A^{+} A^{(n_A)}} \epsilon_t^{(\lambda_1)}\right)  \left( d_u^{A^{(n_A)} A} \epsilon_u^{(\lambda_2)} \right)}{(-\Delta_{1}-\hbar \omega_1) (-\Delta_{A}^{(n_A)}-\hbar \omega_2) (\Delta_{1}+\Delta_{2}+\hbar \omega_1+\hbar \omega_2)}\nonumber \\ &-\frac{e^{-i \vec{p}_1 \cdot \vec{R}_{D_{1} A}+i \vec{p}_2 \cdot \vec{R}_{D_{2} A}} \left( d_r^{A^{+} A^{(n_A)}} \epsilon_r^{(\lambda_1)}\right) \left( d_s^{D_{2} D_{2}^{*}} \epsilon_s^{(\lambda_2)}\right) \left( d_t^{D_{1} D_{1}^{*}} \epsilon_t^{(\lambda_1)}\right)  \left( d_u^{A^{(n_A)} A} \epsilon_u^{(\lambda_2)} \right)}{(\Delta_{1}+\Delta_{2}-\Delta_{A}^{(n_A)}-\hbar \omega_1) (-\Delta_{A}^{(n_A)}-\hbar \omega_2) (-\Delta_{1}+\Delta_{A}^{(n_A)}+\hbar \omega_1+\hbar \omega_2)}\nonumber \\ &-\frac{e^{i \left(\vec{p}_2 \cdot \vec{R}_{D_{1} A}-\vec{p}_1 \cdot \vec{R}_{D_{2} A}\right)} \left( d_r^{A^{+} A^{(n_A)}} \epsilon_r^{(\lambda_1)}\right) \left( d_s^{D_{1} D_{1}^{*}} \epsilon_s^{(\lambda_2)}\right) \left( d_t^{D_{2} D_{2}^{*}} \epsilon_t^{(\lambda_1)}\right)  \left( d_u^{A^{(n_A)} A} \epsilon_u^{(\lambda_2)} \right)}{(\Delta_{1}+\Delta_{2}-\Delta_{A}^{(n_A)}-\hbar \omega_1) (-\Delta_{A}^{(n_A)}-\hbar \omega_2) (-\Delta_{2}+\Delta_{A}^{(n_A)}+\hbar \omega_1+\hbar \omega_2)}\nonumber \\ &+\frac{e^{i \left(\vec{p}_2 \cdot \vec{R}_{D_{1} A}+\vec{p}_1 \cdot \vec{R}_{D_{2} A}\right)} \left( d_r^{D_{2} D_{2}^{*}} \epsilon_r^{(\lambda_1)}\right) \left( d_s^{A^{+} A^{(n_A)}} \epsilon_s^{(\lambda_1)}\right) \left( d_t^{D_{1} D_{1}^{*}} \epsilon_t^{(\lambda_2)}\right)  \left( d_u^{A^{(n_A)} A} \epsilon_u^{(\lambda_2)} \right)}{(\Delta_{1}-\Delta_{A}^{(n_A)}) (-\Delta_{2}-\hbar \omega_1) (-\Delta_{A}^{(n_A)}-\hbar \omega_2)}\nonumber \\ &+\frac{e^{i \left(\vec{p}_2 \cdot \vec{R}_{D_{1} A}-\vec{p}_1 \cdot \vec{R}_{D_{2} A}\right)} \left(d_r^{A^{+} A^{(n_A)}} \epsilon_r^{(\lambda_1)}\right) \left( d_s^{D_{2} D_{2}^{*}} \epsilon_s^{(\lambda_1)}\right) \left( d_t^{D_{1} D_{1}^{*}} \epsilon_t^{(\lambda_2)}\right)  \left( d_u^{A^{(n_A)} A} \epsilon_u^{(\lambda_2)} \right)}{(\Delta_{1}-\Delta_{A}^{(n_A)}) (\Delta_{1}+\Delta_{2}-\Delta_{A}^{(n_A)}-\hbar \omega_1) (-\Delta_{A}^{(n_A)}-\hbar \omega_2)}\nonumber \\ &+\frac{e^{i\left(\vec{p}_1 \cdot \vec{R}_{D_{1} A}+\vec{p}_2 \cdot \vec{R}_{D_{2} A}\right)} \left( d_r^{D_{1} D_{1}^{*}} \epsilon_r^{(\lambda_1)}\right)\left( d_s^{A^{+} A^{(n_A)}} \epsilon_s^{(\lambda_1)}\right) \left( d_t^{D_{2} D_{2}^{*}} \epsilon_t^{(\lambda_2)}\right)  \left( d_u^{A^{(n_A)} A} \epsilon_u^{(\lambda_2)} \right)}{(\Delta_{2}-\Delta_{A}^{(n_A)}) (-\Delta_{1}-\hbar \omega_1) (-\Delta_{A}^{(n_A)}-\hbar \omega_2)}\nonumber \\ &+\frac{e^{-i \vec{p}_1 \cdot \vec{R}_{D_{1} A}+i \vec{p}_2 \cdot \vec{R}_{D_{2} A}} \left( \mu_r^{A^{+} A^{(n_A)}} \epsilon_r^{(\lambda_1)}\right) \left( d_s^{D_{1} D_{1}^{*}} \epsilon_s^{(\lambda_1)}\right) \left( d_t^{D_{2} D_{2}^{*}} \epsilon_t^{(\lambda_2)}\right)  \left( d_u^{A^{(n_A)} A} \epsilon_u^{(\lambda_2)} \right)}{(\Delta_{2}-\Delta_{A}^{(n_A)}) (\Delta_{1}+\Delta_{2}-\Delta_{A}^{(n_A)}-\hbar \omega_1) (-\Delta_{A}^{(n_A)}-\hbar \omega_2)}\nonumber \\ &-\frac{e^{i\left(-\vec{p}_2 \cdot \vec{R}_{D_{1} A}+\vec{p}_1 \cdot \vec{R}_{D_{2} A}\right)} \left( d_r^{D_{2} D_{2}^{*}} \epsilon_r^{(\lambda_1)}\right)\left( d_s^{A^{+} A^{(n_A)}} \epsilon_s^{(\lambda_2)}\right) \left( d_t^{A^{(n_A)} A} \epsilon_t^{(\lambda_1)}\right)  \left( d_u^{D_{1} D_{1}^{*}} \epsilon_u^{(\lambda_2)} \right)}{(-\Delta_{2}-\hbar \omega_1) (\Delta_{1}-\hbar \omega_2) (-\Delta_{1}+\Delta_{A}^{(n_A)}+\hbar \omega_1+\hbar \omega_2)}\nonumber \\ &-\frac{e^{-i \left(\vec{p}_2 \cdot \vec{R}_{D_{1} A}+\vec{p}_1 \cdot \vec{R}_{D_{2} A}\right)} \left( d_r^{A^{+} A^{(n_A)}} \epsilon_r^{(\lambda_1)}\right) \left( d_s^{A^{(n_A)} A} \epsilon_s^{(\lambda_2)}\right) \left( d_t^{D_{2} D_{2}^{*}} \epsilon_t^{(\lambda_1)}\right)  \left( d_u^{D_{1} D_{1}^{*}}\epsilon_u^{(\lambda_2)} \right)}{(\Delta_{1}+\Delta_{2}-\Delta_{A}^{(n_A)}-\hbar \omega_1) (\Delta_{1}-\hbar \omega_2) (-\Delta_{1}-\Delta_{2}+\hbar \omega_1+\hbar \omega_2)}\nonumber \\ &+\frac{e^{i \left(-\vec{p}_2 \cdot \vec{R}_{D_{1} A}+\vec{p}_1 \cdot \vec{R}_{D_{2} A}\right)} \left( d_r^{D_{2} D_{2}^{*}} \epsilon_r^{(\lambda_1)}\right) \left( d_s^{A^{+} A^{(n_A)}}\epsilon_s^{(\lambda_1)}\right) \left( d_t^{A^{(n_A)} A} \epsilon_t^{(\lambda_2)}\right)  \left( d_u^{D_{1} D_{1}^{*}}\epsilon_u^{(\lambda_2)} \right)}{(\Delta_{1}-\Delta_{A}^{(n_A)}) (-\Delta_{2}-\hbar \omega_1) (\Delta_{1}-\hbar \omega_2)}\nonumber \\ &+\frac{e^{-i \left(\vec{p}_2 \cdot \vec{R}_{D_{1} A}+\vec{p}_1 \cdot \vec{R}_{D_{2} A}\right)} \left( d_r^{A^{+} A^{(n_A)}} \epsilon_r^{(\lambda_1)}\right) \left( d_s^{D_{2} D_{2}^{*}} \epsilon_s^{(\lambda_1)}\right) \left( d_t^{A^{(n_A)} A} \epsilon_t^{(\lambda_2)}\right)  \left( d_u^{D_{1} D_{1}^{*}} \epsilon_u^{(\lambda_2)} \right)}{(\Delta_{1}-\Delta_{A}^{(n_A)}) (\Delta_{1}+\Delta_{2}-\Delta_{A}^{(n_A)}-\hbar \omega_1) (\Delta_{1}-\hbar \omega_2)}\nonumber \\ &-\frac{e^{i \left(\vec{p}_1 \cdot \vec{R}_{D_{1} A}-\vec{p}_2 \cdot \vec{R}_{D_{2} A}\right)}\left( d_r^{D_{1} D_{1}^{*}} \epsilon_r^{(\lambda_1)}\right) \left( d_s^{A^{+} A^{(n_A)}} \epsilon_s^{(\lambda_2)}\right) \left( d_t^{A^{(n_A)} A} \epsilon_t^{(\lambda_1)}\right)  \left( d_u^{D_{2} D_{2}^{*}} \epsilon_u^{(\lambda_2)} \right)}{(-\Delta_{1}-\hbar \omega_1) (\Delta_{2}-\hbar \omega_2) (-\Delta_{2}+\Delta_{A}^{(n_A)}+\hbar \omega_1+\hbar \omega_2)}\nonumber \\ &-\frac{e^{-i \left(\vec{p}_1 \cdot \vec{R}_{D_{1} A}+\vec{p}_2 \cdot \vec{R}_{D_{2} A}\right)} \left( d_r^{A^{+} A^{(n_A)}}\epsilon_r^{(\lambda_1)}\right) \left( d_s^{A^{(n_A)} A} \epsilon_s^{(\lambda_2)}\right) \left( d_t^{D_{1} D_{1}^{*}}\epsilon_t^{(\lambda_1)}\right)  \left( d_u^{D_{2} D_{2}^{*}} \epsilon_u^{(\lambda_2)} \right)}{(\Delta_{1}+\Delta_{2}-\Delta_{A}^{(n_A)}-\hbar \omega_1) (\Delta_{2}-\hbar \omega_2) (-\Delta_{1}-\Delta_{2}+\hbar \omega_1+\hbar \omega_2)}\nonumber \\ &+\frac{e^{i \left(\vec{p}_1 \cdot \vec{R}_{D_{1} A}-\vec{p}_2 \cdot \vec{R}_{D_{2} A}\right)} \left( d_r^{D_{1} D_{1}^{*}} \epsilon_r^{(\lambda_1)}\right) \left( d_s^{A^{+} A^{(n_A)}} \epsilon_s^{(\lambda_1)}\right) \left( d_t^{A^{(n_A)} A} \epsilon_t^{(\lambda_2)}\right)  \left( d_u^{D_{2} D_{2}^{*}} \epsilon_u^{(\lambda_2)} \right)}{(\Delta_{2}-\Delta_{A}^{(n_A)}) (-\Delta_{1}-\hbar \omega_1) (\Delta_{2}-\hbar \omega_2)}\nonumber \\ & \left.+\frac{e^{-i \left(\vec{p}_1 \cdot \vec{R}_{D_{1} A}+\vec{p}_2 \cdot \vec{R}_{D_{2} A}\right)} \left(d_r^{A^{+} A^{(n_A)}} \epsilon_r^{(\lambda_1)}\right) \left( d_s^{D_{1} D_{1}^{*}} \epsilon_s^{(\lambda_1)}\right) \left( d_t^{A^{(n_A)} A} \epsilon_t^{(\lambda_2)}\right)  \left( d_u^{D_{2} D_{2}^{*}} \epsilon_u^{(\lambda_2)} \right)}{(\Delta_{2}-\Delta_{A}^{(n_A)}) (\Delta_{1}+\Delta_{2}-\Delta_{A}^{(n_A)}-\hbar \omega_1) (\Delta_{2}-\hbar \omega_2)}\right]
\label{appendix_A_eq_13}
\end{align}
Notice that in Eq. \eqref{appendix_A_eq_13}, we have eight different types of numerators. Four of them are due to the sign in the exponential factors (namely, $e^{\pm i \vec{p}_1 \cdot \vec{R}_{D_{1} A} \pm i \vec{p}_2 \cdot \vec{R}_{D_{2} A}}$) times two different Cartesian indexations of the transition dipole moments of the acceptors, which are $d_x^{A^{(n_A)} A} d_y^{A^{+} A^{(n_A)}}$ and $d_y^{A^{(n_A)} A} d_x^{A^{+} A^{(n_A)}}$ (for instance, $x, y = r, s, t, u$ and $x \neq y$).
Collecting these numerators and performing some algebra on the denominators, one obtains
\begin{equation}
    \begin{split}
        & M_{fi}^{(1)} = - \sum_{r} \sum_{s} \sum_{t} \sum_{u}  \left(\frac{\hbar c }{2 \epsilon_0}\right)^2 \sum_{n_A} \Biggl[ \frac{d_r^{D_{1} D_{1}^{*}} d_s^{A^{(n_A)} A} d_t^{A^{+} A^{(n_A)}} d_u^{D_{2} D_{2}^{*}}}{\Delta_A^{(n_A)} - \Delta_1} + \frac{d_r^{D_{1} D_{1}^{*}} d_s^{A^{+} A^{(n_A)}} d_t^{A^{(n_A)} A} d_u^{D_{2} D_{2}^{*}}}{\Delta_A^{(n_A)} - \Delta_2}
        \Biggr] \\
        \times & \left(\frac{1}{V} \sum_{\vec{p}_1}\right) p_1 \left(\frac{1}{V} \sum_{\vec{p}_2}\right) p_2  \left[\sum_{\lambda_1} \epsilon_r^{(\lambda_1)} (\vec{p}_1) \epsilon_s^{(\lambda_1)}(\vec{p}_1) \right] \left[\sum_{\lambda_2} \epsilon_t^{(\lambda_2)} (\vec{p}_2) \epsilon_u^{(\lambda_2)} (\vec{p}_2) \right]\\
        \times\Biggl[& \frac{e^{+i \vec{p}_1 \cdot \vec{R}_{D_{1} A} + i \vec{p}_2 \cdot \vec{R}_{D_{2} A}}}{\left(\Delta_1 +\hbar \omega_1 \right)\left(\Delta_2 +\hbar \omega_2 \right)} - \frac{e^{+i \vec{p}_1 \cdot \vec{R}_{D_{1} A} - i \vec{p}_2 \cdot \vec{R}_{D_{2} A}}}{\left(\Delta_1 +\hbar \omega_1 \right)\left(\Delta_2 -\hbar \omega_2 \right)} - \frac{e^{- i \vec{p}_1 \cdot \vec{R}_{D_{1} A} + i \vec{p}_2 \cdot \vec{R}_{D_{2} A}}}{\left(\Delta_1 -\hbar \omega_1 \right)\left(\Delta_2 +\hbar \omega_2 \right)} + 
        \frac{e^{-i \vec{p}_1 \cdot \vec{R}_{D_{1} A} - i \vec{p}_2 \cdot \vec{R}_{D_{2} A}}}{\left(\Delta_1 -\hbar \omega_1 \right)\left(\Delta_2 -\hbar \omega_2 \right)}\Biggr]
    \end{split}
    \label{appendix_A_eq_14}
\end{equation}
where $\vec{\epsilon}^{(\lambda_1)} = \vec{\epsilon}^{(\lambda_1)} (\vec{p}_1)$ and $\vec{\epsilon}^{(\lambda_2)} = \vec{\epsilon}^{(\lambda_2)} (\vec{p}_2)$.
To proceed, we use the following relations for the polarization vectors and the transformation of the summation in the virtual photon momenta summation to an integral representation \cite{craig1998molecular}
\begin{equation}
    \sum_{\lambda} \epsilon_i^{(\lambda)} (\vec{p}) \epsilon_j^{(\lambda_1)}(\vec{p}) = \delta_{ij} - \hat{p}_i \hat{p}_j, \qquad 
    \frac{1}{V} \sum_{\vec{p}} \rightarrow \frac{1}{(2 \pi)^3} \int d^3 p
    \label{appendix_A_eq_15}
\end{equation}
After some algebra in the third line of Eq. \eqref{appendix_A_eq_14} and using Eqs. \eqref{terms_eq_9} and \eqref{appendix_A_eq_15}, we get
\begin{equation}
    \begin{split}
        M_{fi}^{(1)} = - & \sum_{r} \sum_{s} \sum_{t} \sum_{u}  d_r^{D_{1} D_{1}^{*}} \alpha_{st}^{A} (k_1, k_2) d_u^{D_{2} D_{2}^{*}} \\
        \times & \left(\frac{1}{2 \epsilon_0} \right) \int \frac{d^3 p_1}{(2 \pi)^3} p_1^2 \left( \delta_{rs} - \hat{p}_{1,r} \hat{p}_{1,s} \right) \left(\frac{e^{-i \vec{p}_1 \cdot \vec{R}_{D_{1} A}}}{k_1 - p_1} - \frac{e^{+i \vec{p}_1 \cdot \vec{R}_{D_{1} A}}}{k_1 + p_1}\right) \\
        \times & \left(\frac{1}{2 \epsilon_0}\right) \int \frac{d^3 p_{2}}{(2 \pi)^3} p_2^2 \left( \delta_{tu} - \hat{p}_{2,t} \hat{p}_{2,u} \right)  \left(\frac{e^{-i \vec{p}_2 \cdot \vec{R}_{D_{2} A}}}{k_2 - p_2} - \frac{e^{+i \vec{p}_2 \cdot \vec{R}_{D_{2} A}}}{k_2 + p_2}\right)
    \end{split}
    \label{appendix_A_eq_16}
\end{equation}
where we used $\hbar \omega_1 = \hbar c p_1$ and $\hbar \omega_2 = \hbar c p_2$, as well as $\Delta_A^{(n_A)} = \hbar c k_A^{(n_A)}$, $\Delta_1 = \hbar c k_1$ and $\Delta_2 = \hbar c k_2$.
The integrals in Eq. \eqref{appendix_A_eq_16} are formally identical and their result is given by the retarded interaction tensor in Eq. \eqref{terms_eq_12} as detailed in Ref. \onlinecite{daniels2003resonance}. It follows that
\begin{equation}
    \begin{split}
        & \left(\frac{1}{2 \epsilon_0} \right) \int \frac{d^3 p_1}{(2 \pi)^3} p_1^2 \left( \delta_{rs} - \hat{p}_{1,r} \hat{p}_{1,s} \right) \left(\frac{e^{-i \vec{p}_1 \cdot \vec{R}_{D_{1} A}}}{k_1 - p_1} - \frac{e^{+i \vec{p}_1 \cdot \vec{R}_{D_{1} A}}}{k_1 + p_1}\right) = V_{rs} (k_1, \vec{R}_{D_1 A}) \\
        & \left(\frac{1}{2 \epsilon_0}\right) \int \frac{d^3 p_{2}}{(2 \pi)^3} p_2^2 \left( \delta_{tu} - \hat{p}_{2,t} \hat{p}_{2,u} \right)  \left(\frac{e^{-i \vec{p}_2 \cdot \vec{R}_{D_{2} A}}}{k_2 - p_2} - \frac{e^{+i \vec{p}_2 \cdot \vec{R}_{D_{2} A}}}{k_2 + p_2}\right) = V_{tu} (k_2, \vec{R}_{D_2 A}) \\
    \end{split}
    \label{appendix_A_eq_17}
\end{equation}
By inserting Eq. \eqref{appendix_A_eq_17} into Eq. \eqref{appendix_A_eq_16}, we finally obtain:
\begin{equation}
        M_{fi}^{(1)} = - \sum_{r} \sum_{s} \sum_{t} \sum_{u} d_r^{D_{1} D_{1}^{*}} V_{rs} (k_1, \vec{R}_{D_1 A}) \alpha_{st}^{A} (k_1, k_2) V_{tu} (k_2, \vec{R}_{D_2 A}) d_u^{D_{2} D_{2}^{*}} 
    \label{appendix_A_eq_18}
\end{equation}
where the result in Eq. \eqref{appendix_A_eq_18} is exactly the same shown in Eq. \eqref{terms_eq_2}.

\section{Amplitudes from the diagrammatic representation employing the two-body interactions}
\label{appendix_B}
From Figure \ref{methods_fig_2} and using the procedures introduced in Subsection \ref{diagrammatic_representation}, the transition amplitudes read (the superscript of an amplitude refers to the diagrams (a) to (f) in Figure 2)
\begin{equation}
    M_{fi}^{(a)} = \frac{\langle D_1^{*} , D_2^{*}| \hat{W} (k_1, \vec{R}_{D_1 D_2}) | D_1 , D_2^{(j)} \rangle \langle D_2^{(j)} , A | \hat{W} (k_1 + k_2, \vec{R}_{D_2 A}) | D_2 , A \rangle}{(E_{D_1^*} + E_{D_2^{*}} + E_{A}) - (E_{D_1} + E_{D_2^{(j)}} + E_{A})}
    \label{appendix_B_eq_1}
\end{equation}
\begin{equation}
    M_{fi}^{(b)} = \frac{\langle D_2^{*} , A| \hat{W} (k_1 + k_2, \vec{R}_{D_2 A}) | D_2^{(j)} , A^{+} \rangle \langle D_1^{*}, D_2^{(j)} | \hat{W} (k_1, \vec{R}_{D_1 D_2}) | D_1, D_2 \rangle}{(E_{D_1^*} + E_{D_2^{*}} + E_{A}) - (E_{D_1^{*}} + E_{D_2^{(j)}} + E_{A^{+}})}
    \label{appendix_B_eq_2}
\end{equation}
\begin{equation}
    M_{fi}^{(c)} = \frac{\langle D_1^{*} , D_2^{*}| \hat{W} (k_2, \vec{R}_{D_1 D_2}) | D_1^{(j)} , D_2 \rangle \langle D_1^{(j)} , A | \hat{W} (k_1 + k_2, \vec{R}_{D_1 A}) | D_1 , A \rangle}{(E_{D_1^*} + E_{D_2^{*}} + E_{A}) - (E_{D_1^{(j)}} + E_{D_2} + E_{A})}
    \label{appendix_B_eq_3}
\end{equation}
\begin{equation}
    M_{fi}^{(d)} = \frac{\langle D_1^{*} , A| \hat{W} (k_1 + k_2, \vec{R}_{D_1 A}) | D_1^{(j)} , A^{+} \rangle \langle D_1^{(j)}, D_2^{*} | \hat{W} (k_2, \vec{R}_{D_1 D_2}) | D_1, D_2 \rangle}{(E_{D_1^*} + E_{D_2^{*}} + E_{A}) - (E_{D_1^{(j)}} + E_{D_2^{*}} + E_{A^{+}})}
    \label{appendix_B_eq_4}
\end{equation}
\begin{equation}
    M_{fi}^{(e)} = \frac{\langle D_1^{*} , A| \hat{W} (k_1, \vec{R}_{D_1 A}) | D_1 , A^{(n_A)} \rangle \langle D_2^{*}, A^{(n_A)} | \hat{W} (k_2, \vec{R}_{D_2 A}) | D_2, A^{+} \rangle}{(E_{D_1^*} + E_{D_2^{*}} + E_{A}) - (E_{D_1} + E_{D_2^{*}} + E_{A^{(n_A)}})}
    \label{appendix_B_eq_5}
\end{equation}
\begin{equation}
    M_{fi}^{(f)} = \frac{\langle D_2^{*} , A| \hat{W} (k_2, \vec{R}_{D_2 A}) | D_2 , A^{(n_A)} \rangle \langle D_1^{*}, A^{(n_A)} | \hat{W} (k_1, \vec{R}_{D_1 A}) | D_1, A^{+} \rangle}{(E_{D_1^*} + E_{D_2^{*}} + E_{A}) - (E_{D_1}^{*} + E_{D_2} + E_{A^{(n_A)}})}
    \label{appendix_B_eq_6}
\end{equation}
The correspondence of the amplitudes in Eqs. \eqref{appendix_B_eq_1} -- \eqref{appendix_B_eq_6} and those in Eqs. \eqref{terms_eq_2}--\eqref{terms_eq_8} is
\begin{equation}
    \sum_{n_A} \left( M_{fi}^{(e)} + M_{fi}^{(f)} \right) = M_{fi}^{(1)}
    \label{appendix_B_eq_7}
\end{equation}
\begin{equation}
    M_{fi}^{(c)} \left\{D_1 \right\} + M_{fi}^{(d)} \left\{D_1^{*} \right\} = M_{fi}^{(2)}
    \label{appendix_B_eq_8}
\end{equation}
\begin{equation}
    M_{fi}^{(c)} \left\{D_1^{*} \right\} + M_{fi}^{(d)} \left\{D_1 \right\} = M_{fi}^{(3)}
    \label{appendix_B_eq_9}
\end{equation}
\begin{equation}
    M_{fi}^{(a)} \left\{D_2 \right\} + M_{fi}^{(b)} \left\{D_2^{*} \right\} = M_{fi}^{(4)}
    \label{appendix_B_eq_10}
\end{equation}
\begin{equation}
    M_{fi}^{(a)} \left\{D_2^{*} \right\} + M_{fi}^{(b)} \left\{D_2 \right\} = M_{fi}^{(5)}
    \label{appendix_B_eq_11}
\end{equation}
\begin{equation}
   \sum_{n_1} \left(M_{fi}^{(c)} \left\{D_1^{(n_1)} \right\} + M_{fi}^{(d)} \left\{ D_1^{(n_1)} \right\} \right) = M_{fi}^{(6)}
    \label{appendix_B_eq_12}
\end{equation}
\begin{equation}
   \sum_{n_2} \left(M_{fi}^{(a)} \left\{ D_2^{(n_2)} \right\} + M_{fi}^{(b)} \left\{D_2^{(n_2)} \right\} \right) = M_{fi}^{(7)}
    \label{appendix_B_eq_13}
\end{equation}

\section{Terms of collective ICD rate averaged over orientations}
\label{appendix_C}
In this Appendix, we will continue to discuss the technique of averaging over orientations of the collective ICD rate and list the relevant terms. We concentrate on the terms that include the elements $\vec{d}^{\delta D_1}$ and $\vec{d}^{\delta D_2}$. 
Notice that for the rates associated with transition amplitudes which include the transition polarizabilities, namely, $M_{fi}^{(1)}, M_{fi}^{(6)}, M_{fi}^{(7)}$, the procedures discussed both in Subsection \ref{averaging_over_orientations} and below are also valid.

From Eq. \eqref{avg_eq_3}, the total rate (associated to $\vec{d}^{\delta D_1}$ and $\vec{d}^{\delta D_2}$) is given by
\begin{equation}
    \langle \Gamma \rangle = \langle \Gamma^{(2,2)} \rangle + \langle \Gamma^{(3,3)} \rangle + \langle \Gamma^{(4,4)} \rangle + \langle \Gamma^{(5,5)} \rangle + 2 \Re \langle \Gamma^{(2,3)} \rangle + 2 \Re \langle \Gamma^{(4,5)} \rangle
    \label{appendix_C_eq_0}
\end{equation}
The transition amplitudes related to the partial rates in Eq. \eqref{appendix_C_eq_0} are collected in Eqs. \eqref{terms_eq_3}--\eqref{terms_eq_6}. 
Following the same averaging procedure employed in Eqs. \eqref{avg_eq_11} -- \eqref{avg_eq_14}, we find
\begin{equation}
    \begin{split}
         \langle \Gamma^{(2,2)} \rangle = & \frac{2 \pi }{270 \hbar (\Delta_1 + \Delta_2)^2}
         \sum_{r,s,t,u}  \sum_{r',s',t',u'}
        \lvert \vec{d}^{D_2 D_2^{*}} \rvert^2 \lvert \vec{d}^{A^{+} A} \rvert^2 \delta_{s s'} \delta_{t t'} \\
          & \times \Biggl\{ \left[4 \left( \vec{d}^{\delta D_1} \cdot \vec{d}^{D_1 D_1^{*}}\right)^2 - 2 \lvert \vec{d}^{\delta D_1} \rvert^2 \lvert \vec{d}^{D_1 D_1^{*}}  \rvert^2 \right]  \delta_{r u}  \delta_{r' u'} \\
        & - \left[\left( \vec{d}^{\delta D_1} \cdot \vec{d}^{D_1 D_1^{*}}\right)^2 - 3 \lvert \vec{d}^{\delta D_1} \rvert^2 \lvert \vec{d}^{D_1 D_1^{*}}  \rvert^2 \right] \left(\delta_{r u'} \delta_{u r'} + \delta_{r r'} \delta_{u u'} \right) \Biggr\}\\
         &\times V_{rs} (k_1 + k_2, \vec{R}_{D_{1}A}) V_{r's'}^{*} (k_1 + k_2, \vec{R}_{D_{1}A}) V_{tu} (k_2, \vec{R}_{D_{1}D_{2}}) V_{t'u'}^{*} (k_2, \vec{R}_{D_{1}D_{2}}) 
    \end{split}
    \label{appendix_C_eq_1}
\end{equation}
\begin{equation}
    \begin{split}
         \langle \Gamma^{(3,3)} \rangle = & \frac{2 \pi }{270 \hbar (\Delta_2)^2}
         \sum_{r,s,t,u}  \sum_{r',s',t',u'}
        \lvert \vec{d}^{D_2 D_2^{*}} \rvert^2 \lvert \vec{d}^{A^{+} A} \rvert^2 \delta_{s s'} \delta_{t t'} \\
         & \times \Biggl\{ \left[4 \left( \vec{d}^{\delta D_1} \cdot \vec{d}^{D_1 D_1^{*}}\right)^2 - 2 \lvert \vec{d}^{\delta D_1} \rvert^2 \lvert \vec{d}^{D_1 D_1^{*}}  \rvert^2 \right]  \delta_{r u}  \delta_{r' u'} \\
        & - \left[\left( \vec{d}^{\delta D_1} \cdot \vec{d}^{D_1 D_1^{*}}\right)^2 - 3 \lvert \vec{d}^{\delta D_1} \rvert^2 \lvert \vec{d}^{D_1 D_1^{*}}  \rvert^2 \right] \left(\delta_{r u'} \delta_{u r'} + \delta_{r r'} \delta_{u u'} \right) \Biggr\}\\
        & \times V_{rs} (k_1 + k_2, \vec{R}_{D_{1}A}) V_{r's'}^{*} (k_1 + k_2, \vec{R}_{D_{1}A}) V_{tu} (k_2, \vec{R}_{D_{1}D_{2}}) V_{t'u'}^{*} (k_2, \vec{R}_{D_{1}D_{2}}) 
    \end{split}
    \label{appendix_C_eq_2}
\end{equation}
\begin{equation}
    \begin{split}
         \langle \Gamma^{(4,4)} \rangle = & \frac{2 \pi }{270 \hbar (\Delta_1 + \Delta_2)^2} \sum_{r,s,t,u}  \sum_{r',s',t',u'}
        \lvert \vec{d}^{D_1 D_1^{*}} \rvert^2 \lvert \vec{d}^{A^{+} A} \rvert^2 \delta_{s s'} \delta_{t t'} \\
          & \times \Biggl\{ \left[4 \left( \vec{d}^{\delta D_2} \cdot \vec{d}^{D_2 D_2^{*}}\right)^2 - 2 \lvert \vec{d}^{\delta D_2} \rvert^2 \lvert \vec{d}^{D_2 D_2^{*}}  \rvert^2 \right]  \delta_{r u}  \delta_{r' u'} \\
         & - \left[\left( \vec{d}^{\delta D_2} \cdot \vec{d}^{D_2 D_2^{*}}\right)^2 - 3 \lvert \vec{d}^{\delta D_2} \rvert^2 \lvert \vec{d}^{D_2 D_2^{*}}  \rvert^2 \right] \left(\delta_{r u'} \delta_{u r'} + \delta_{r r'} \delta_{u u'} \right) \Biggr\}\\
        & \times V_{rs} (k_1 + k_2, \vec{R}_{D_{2}A}) V_{r's'}^{*} (k_1 + k_2, \vec{R}_{D_{2}A}) V_{tu} (k_1, \vec{R}_{D_{1}D_{2}}) V_{t'u'}^{*} (k_1, \vec{R}_{D_{1}D_{2}}) 
    \end{split}
    \label{appendix_C_eq_3}
\end{equation}
\begin{equation}
    \begin{split}
         \langle \Gamma^{(5,5)} \rangle = & \frac{2 \pi }{270 \hbar (\Delta_1)^2}
         \sum_{r,s,t,u}  \sum_{r',s',t',u'}
        \lvert \vec{d}^{D_1 D_1^{*}} \rvert^2 \lvert \vec{d}^{A^{+} A} \rvert^2 \delta_{s s'} \delta_{t t'} \\
         & \times \Biggl\{ \left[4 \left( \vec{d}^{\delta D_2} \cdot \vec{d}^{D_2 D_2^{*}}\right)^2 - 2 \lvert \vec{d}^{\delta D_2} \rvert^2 \lvert \vec{d}^{D_2 D_2^{*}}  \rvert^2 \right]  \delta_{r u}  \delta_{r' u'} \\
        & - \left[\left( \vec{d}^{\delta D_2} \cdot \vec{d}^{D_2 D_2^{*}}\right)^2 - 3 \lvert \vec{d}^{\delta D_2} \rvert^2 \lvert \vec{d}^{D_2 D_2^{*}}  \rvert^2 \right] \left(\delta_{r u'} \delta_{u r'} + \delta_{r r'} \delta_{u u'} \right) \Biggr\}\\
        &\times V_{rs} (k_1 + k_2, \vec{R}_{D_{2}A}) V_{r's'}^{*} (k_1 + k_2, \vec{R}_{D_{2}A}) V_{tu} (k_1, \vec{R}_{D_{1}D_{2}}) V_{t'u'}^{*} (k_1, \vec{R}_{D_{1}D_{2}}) 
    \end{split}
    \label{appendix_C_eq_4}
\end{equation}
\begin{equation}
    \begin{split}
         \langle \Gamma^{(2,3)} \rangle = &  -\frac{2 \pi }{270 \hbar (\Delta_2)(\Delta_1 + \Delta_2)}
         \sum_{r,s,t,u}  \sum_{r',s',t',u'}
        \lvert \vec{d}^{D_2 D_2^{*}} \rvert^2 \lvert \vec{d}^{A^{+} A} \rvert^2 \delta_{s s'} \delta_{t t'} \\
         & \times \Biggl\{ \left[4 \lvert \vec{d}^{\delta D_1} \rvert^2 \lvert \vec{d}^{D_1 D_1^{*}}  \rvert^2 - 2 \left( \vec{d}^{\delta D_1} \cdot \vec{d}^{D_1 D_1^{*}}\right)^2 \right]  \delta_{r u}  \delta_{r' u'} \\
         & - \left[\lvert \vec{d}^{\delta D_1} \rvert^2 \lvert \vec{d}^{D_1 D_1^{*}}  \rvert^2 - 3 \left( \vec{d}^{\delta D_1} \cdot \vec{d}^{D_1 D_1^{*}}\right)^2 \right] \left(\delta_{r u'} \delta_{u r'} + \delta_{r r'} \delta_{u u'} \right) \Biggr\}\\
        & \times V_{rs} (k_1 + k_2, \vec{R}_{D_{1}A}) V_{r's'}^{*} (k_1 + k_2, \vec{R}_{D_{1}A}) V_{tu} (k_2, \vec{R}_{D_{1}D_{2}}) V_{t'u'}^{*} (k_2, \vec{R}_{D_{1}D_{2}}) 
    \end{split}
    \label{appendix_C_eq_5}
\end{equation}
\begin{equation}
    \begin{split}
         \langle \Gamma^{(4,5)} \rangle = & -\frac{2 \pi }{270 \hbar (\Delta_1)(\Delta_1 + \Delta_2)}
         \sum_{r,s,t,u}  \sum_{r',s',t',u'}
        \lvert \vec{d}^{D_1 D_1^{*}} \rvert^2 \lvert \vec{d}^{A^{+} A} \rvert^2 \delta_{s s'} \delta_{t t'} \\
         & \Biggl\{ \left[4 \lvert \vec{d}^{\delta D_2} \rvert^2 \lvert \vec{d}^{D_2 D_2^{*}}  \rvert^2 - 2 \left( \vec{d}^{\delta D_2} \cdot \vec{d}^{D_2 D_2^{*}}\right)^2 \right]  \delta_{r u}  \delta_{r' u'} \\
        & - \left[\lvert \vec{d}^{\delta D_2} \rvert^2 \lvert \vec{d}^{D_2 D_2^{*}}  \rvert^2 - 3 \left( \vec{d}^{\delta D_2} \cdot \vec{d}^{D_2 D_2^{*}}\right)^2 \right] \left(\delta_{r u'} \delta_{u r'} + \delta_{r r'} \delta_{u u'} \right) \Biggr\}\\
        &\times V_{rs} (k_1 + k_2, \vec{R}_{D_{2}A}) V_{r's'}^{*} (k_1 + k_2, \vec{R}_{D_{2}A}) V_{tu} (k_1, \vec{R}_{D_{1}D_{2}}) V_{t'u'}^{*} (k_1, \vec{R}_{D_{1}D_{2}}) 
    \end{split}
    \label{appendix_C_eq_6}
\end{equation}

For all rates in Eqs. \eqref{appendix_C_eq_1}--\eqref{appendix_C_eq_6}, we need to evaluate the following three entities:
\begin{equation}
    \begin{split}
        \mathcal{S}_1 (k_{\ell'}, \vec{R}_{D_{\ell}}) = \sum_{r,s,t,u}  \sum_{r',s',t',u'} & V_{rs} (k_1 + k_2, \vec{R}_{D_{\ell}A}) V_{r's'}^{*} (k_1 + k_2, \vec{R}_{{D}_{\ell}A}) V_{tu} (k, \vec{R}_{D_{1}D_{2}}) V_{t'u'}^{*} (k_{\ell'}, \vec{R}_{D_{1}D_{2}}) \\
        & \times \delta_{s s'} \delta_{t t'} \delta_{r u}  \delta_{r' u'}
    \end{split}
    \label{appendix_C_eq_7}
\end{equation}
\begin{equation}
    \begin{split}
        \mathcal{S}_2 (k_{\ell'}, \vec{R}_{D_{\ell}}) = \sum_{r,s,t,u}  \sum_{r',s',t',u'} & V_{rs} (k_1 + k_2, \vec{R}_{D_{\ell}A}) V_{r's'}^{*} (k_1 + k_2, \vec{R}_{D_{\ell}A}) V_{tu} (k_{\ell'}, \vec{R}_{D_{1}D_{2}}) V_{t'u'}^{*} (k_{\ell'}, \vec{R}_{D_{1}D_{2}}) \\
        & \times \delta_{s s'} \delta_{t t'} \delta_{r u'} \delta_{u r'}
    \end{split}
    \label{appendix_C_eq_8}
\end{equation}
\begin{equation}
    \begin{split}
        \mathcal{S}_3 (k_{\ell'}, \vec{R}_{D_{\ell}}) = \sum_{r,s,t,u}  \sum_{r',s',t',u'} & V_{rs} (k_1 + k_2, \vec{R}_{D_{\ell}A}) V_{r's'}^{*} (k_1 + k_2, \vec{R}_{D_{\ell}A}) V_{tu} (k_{\ell'}, \vec{R}_{D_{1}D_{2}}) V_{t'u'}^{*} (k_{\ell'}, \vec{R}_{D_{1}D_{2}}) \\
        & \times \delta_{s s'} \delta_{t t'} \delta_{r r'} \delta_{u u'}
    \end{split}
    \label{appendix_C_eq_9}
\end{equation}
where $D_{\ell}=D_1, D_2$ and $k_{\ell'} = k_1, k_2$.
By applying the Kronecker deltas and renaming indices, we obtain
\begin{equation}
    \mathcal{S}_1 (\ell, \ell')  = \sum_{r,s,t,u}  V_{rs} (k_1 + k_2, \vec{R}_{D_{\ell}A}) V_{us}^{*} (k_1 + k_2, \vec{R}_{{D}_{\ell}A}) V_{tr} (k_{\ell'}, \vec{R}_{D_{1}D_{2}}) V_{tu}^{*} (k_{\ell'}, \vec{R}_{D_{1}D_{2}}) 
    \label{appendix_C_eq_10}
\end{equation}
\begin{equation}
    \mathcal{S}_2 (\ell, \ell') = \sum_{r,s,t,u}  V_{rs} (k_1 + k_2, \vec{R}_{D_{\ell}A}) V_{us}^{*} (k_1 + k_2, \vec{R}_{D_{\ell}A}) V_{tu} (k_{\ell'}, \vec{R}_{D_{1}D_{2}}) V_{tr}^{*} (k_{\ell'}, \vec{R}_{D_{1}D_{2}}) 
    \label{appendix_C_eq_11}
\end{equation}
\begin{equation}
    \mathcal{S}_3 (\ell, \ell') = \sum_{r,s,t,u}  V_{rs} (k_1 + k_2, \vec{R}_{D_{\ell}A}) V_{rs}^{*} (k_1 + k_2, \vec{R}_{D_{\ell}A}) V_{tu} (k_{\ell'}, \vec{R}_{D_{1}D_{2}}) V_{tu}^{*} (k_{\ell'}, \vec{R}_{D_{1}D_{2}}) 
    \label{appendix_C_eq_12}
\end{equation}
Notice that $\mathcal{S}_1 = \mathcal{S}_2$ since $V_{tr} (k_{\ell'}, \vec{R}_{D_{1}D_{2}}) V_{tu}^{*} (k_{\ell'}, \vec{R}_{D_{1}D_{2}}) = V_{tu} (k_{\ell'}, \vec{R}_{D_{1}D_{2}}) V_{tr}^{*} (k_{\ell'}, \vec{R}_{D_{1}D_{2}})$ (for this, see Eq. \eqref{terms_eq_12}).
Then, using Eqs. \eqref{appendix_C_eq_10}--\eqref{appendix_C_eq_12} in Eqs. \eqref{appendix_C_eq_1}--\eqref{appendix_C_eq_6}, we can rewrite the various contributions to the rates as 
\begin{equation}
    \begin{split}
         \langle \Gamma^{(2,2)} \rangle = &\frac{2 \pi }{270 \hbar (\Delta_1 + \Delta_2)^2}
        \lvert \vec{d}^{D_2 D_2^{*}} \rvert^2 \lvert \vec{d}^{A^{+} A} \rvert^2 \lvert \vec{d}^{\delta D_1} \rvert^2 \lvert \vec{d}^{D_1 D_1^{*}}  \rvert^2 \\
         &  \times \Biggl\{ \left[3 \left( \hat{d}^{\delta D_1} \cdot \hat{d}^{D_1 D_1^{*}}\right)^2 + 1  \right] \mathcal{S}_1 (\ell = 1, \ell' = 2) - \left[\left( \hat{d}^{\delta D_1} \cdot \hat{d}^{D_1 D_1^{*}}\right)^2 - 3 \right] \mathcal{S}_3 (\ell = 1, \ell' = 2)\Biggr\}
    \end{split}
    \label{appendix_C_eq_1x}
\end{equation}
\begin{equation}
    \begin{split}
         \langle \Gamma^{(3,3)} \rangle =  &\frac{2 \pi }{270 \hbar (\Delta_2)^2}
        \lvert \vec{d}^{D_2 D_2^{*}} \rvert^2 \lvert \vec{d}^{A^{+} A} \rvert^2 \lvert \vec{d}^{\delta D_1} \rvert^2 \lvert \vec{d}^{D_1 D_1^{*}}  \rvert^2 \\
         &  \times \Biggl\{ \left[3 \left( \hat{d}^{\delta D_1} \cdot \hat{d}^{D_1 D_1^{*}}\right)^2 + 1  \right] \mathcal{S}_1 (\ell = 1, \ell' = 2) - \left[\left( \hat{d}^{\delta D_1} \cdot \hat{d}^{D_1 D_1^{*}}\right)^2 - 3 \right] \mathcal{S}_3 (\ell = 1, \ell' = 2)\Biggr\}
    \end{split}
    \label{appendix_C_eq_2x}
\end{equation}
\begin{equation}
    \begin{split}
        \langle \Gamma^{(4,4)} \rangle = &\frac{2 \pi }{270 \hbar (\Delta_1 + \Delta_2)^2}
        \lvert \vec{d}^{D_1 D_1^{*}} \rvert^2 \lvert \vec{d}^{A^{+} A} \rvert^2 \lvert \vec{d}^{\delta D_2} \rvert^2 \lvert \vec{d}^{D_2 D_2^{*}}  \rvert^2 \\
        & \times \Biggl\{ \left[3 \left( \hat{d}^{\delta D_2} \cdot \hat{d}^{D_2 D_2^{*}}\right)^2 + 1  \right] \mathcal{S}_1 (\ell = 2, \ell' = 1) - \left[\left( \hat{d}^{\delta D_2} \cdot \hat{d}^{D_2 D_2^{*}}\right)^2 - 3 \right] \mathcal{S}_3 (\ell = 2, \ell' = 1)\Biggr\}
    \end{split}
    \label{appendix_C_eq_3x}
\end{equation}
\begin{equation}
    \begin{split}
        \langle \Gamma^{(5,5)} \rangle = &\frac{2 \pi }{270 \hbar (\Delta_1)^2}
        \lvert \vec{d}^{D_1 D_1^{*}} \rvert^2 \lvert \vec{d}^{A^{+} A} \rvert^2 \lvert \vec{d}^{\delta D_2} \rvert^2 \lvert \vec{d}^{D_2 D_2^{*}}  \rvert^2 \\
         & \times \Biggl\{ \left[3 \left( \hat{d}^{\delta D_2} \cdot \hat{d}^{D_2 D_2^{*}}\right)^2 + 1  \right] \mathcal{S}_1 (\ell = 2, \ell' = 1) - \left[\left( \hat{d}^{\delta D_2} \cdot \hat{d}^{D_2 D_2^{*}}\right)^2 - 3 \right] \mathcal{S}_3 (\ell = 2, \ell' = 1) \Biggr\} 
    \end{split}
    \label{appendix_C_eq_4x}
\end{equation}
\begin{equation}
    \begin{split}
         \langle \Gamma^{(2,3)} \rangle = & - \frac{2 \pi }{270 \hbar \Delta_2 (\Delta_1 + \Delta_2)}
        \lvert \vec{d}^{D_2 D_2^{*}} \rvert^2 \lvert \vec{d}^{A^{+} A} \rvert^2 \lvert \vec{d}^{\delta D_1} \rvert^2 \lvert \vec{d}^{D_1 D_1^{*}}  \rvert^2 \\
          & \times \Biggl\{ \left[3 + \left( \hat{d}^{\delta D_1} \cdot \hat{d}^{D_1 D_1^{*}}\right)^2  \right] \mathcal{S}_1 (\ell = 1, \ell' = 2) - \left[1 - 3 \left( \hat{d}^{\delta D_1} \cdot \hat{d}^{D_1 D_1^{*}}\right)^2 \right] \mathcal{S}_3 (\ell = 1, \ell' = 2)\Biggr\}
    \end{split}
    \label{appendix_C_eq_5x}
\end{equation}
\begin{equation}
    \begin{split}
        \langle \Gamma^{(4,5)} \rangle = & - \frac{2 \pi }{270 \hbar (\Delta_1) (\Delta_1 + \Delta_2)}
        \lvert \vec{d}^{D_1 D_1^{*}} \rvert^2 \lvert \vec{d}^{A^{+} A} \rvert^2 \lvert \vec{d}^{\delta D_2} \rvert^2 \lvert \vec{d}^{D_2 D_2^{*}}  \rvert^2 \\
          & \times \Biggl\{ \left[3 + \left( \hat{d}^{\delta D_2} \cdot \hat{d}^{D_2 D_2^{*}}\right)^2 \right] \mathcal{S}_1 (\ell = 2, \ell' = 1) - \left[1 - 3 \left( \hat{d}^{\delta D_2} \cdot \hat{d}^{D_2 D_2^{*}}\right)^2 \right] \mathcal{S}_3 (\ell = 2, \ell' = 1) \Biggr\} 
    \end{split}
    \label{appendix_C_eq_6x}
\end{equation}

Consequently, to obtain the total rate, our effort now relies on evaluating the expressions for $\mathcal{S}_1$ and $\mathcal{S}_3$ given in Eqs. \eqref{appendix_C_eq_10} and \eqref{appendix_C_eq_12}. 
To do that, we first rewrite the interaction retarded tensor $V_{rs}$ from Eq. \eqref{terms_eq_12} as
\begin{equation}
    V_{rs} (k, \vec{R}) = \mathcal{A} (k, R) \delta_{rs} + \mathcal{B} (k, R)\hat{R}_r \hat{R}_s
    \label{appendix_C_eq_13}
\end{equation}
where the complex auxiliary functions $\mathcal{A}(k, R)$ and $\mathcal{B}(k, R)$ are defined as
\begin{equation}
    \begin{split}
        \mathcal{A} (k, R) = \frac{e^{ikR}}{4 \pi \varepsilon_0 R^3}\left[ 1 - ikR-(kR)^2 \right], \qquad 
        \mathcal{B} (k, R) = \frac{e^{ikR}}{4 \pi \varepsilon_0 R^3}\left[ - 3 + 3ikR+(kR)^2 \right] 
    \end{split}
    \label{appendix_C_eq_14}
\end{equation}
Then, considering first the function $\mathcal{S}_1$, we can conveniently write it as 
\begin{equation}
    \mathcal{S}_1 = \sum_{r,u}  \mathcal{U}_{ru} (k_1 + k_2, \vec{R}_{D_{\ell}A}) \mathcal{W}_{ru} (k_{\ell'}, \vec{R}_{D_{1}D_{2}}) 
    \label{appendix_C_eq_14x}
\end{equation}
where 
\begin{equation}
    \mathcal{W}_{ru} (k_{\ell'}, \vec{R}_{D_{1}D_{2}}) =  \sum_{t} V_{tr} (k_{\ell'}, \vec{R}_{D_{1}D_{2}}) V_{tu}^{*} (k_{\ell'}, \vec{R}_{D_{1}D_{2}})
    \label{appendix_C_eq_15x}
\end{equation}
and
\begin{equation}
     \mathcal{U}_{ru} (k_1 + k_2, \vec{R}_{D_{\ell}A}) =  \sum_{s} V_{rs} (k_1 + k_2, \vec{R}_{D_{\ell}A}) V_{us}^{*} (k_1 + k_2, \vec{R}_{D_{\ell}A})
    \label{appendix_C_eq_16x}
\end{equation}
Inserting Eq. \eqref{appendix_C_eq_13} in Eqs. \eqref{appendix_C_eq_15x} and \eqref{appendix_C_eq_16x} and using $\sum_{t} \hat{R}_t^{D_{1}D_{2}} \hat{R}_t^{D_{1}D_{2}} = 1$, we obtain
\begin{equation}
    \begin{split}
        \mathcal{W}_{ru} (k_{\ell'}, \vec{R}_{D_{1}D_{2}}) =\lvert \mathcal{A} (k_{\ell'}, R_{D_{1}D_{2}}) \rvert^2 \delta_{ru} + \lvert \mathcal{B} (k_{\ell'}, R_{D_{1}D_{2}}) \rvert^2 \hat{R}_r^{D_{1}D_{2}} \hat{R}_u^{D_{1}D_{2}} \\
        + 2 \Re \{ \mathcal{A} (k_{\ell'}, R_{D_{1}D_{2}}) \left[\mathcal{B}(k_{\ell'}, R_{D_{1}D_{2}})\right]^{*}\} \hat{R}_r^{D_{1}D_{2}} \hat{R}_u^{D_{1}D_{2}}
    \end{split}
    \label{appendix_C_eq_15}
\end{equation}
and
\begin{equation}
    \begin{split}
        \mathcal{U}_{ru} (k_1 + k_2, \vec{R}_{D_{\ell}A}) =  \lvert \mathcal{A} (k_1 + k_2, R_{D_{\ell}A}) \rvert^2 \delta_{ru} + \lvert \mathcal{B} (k_1 + k_2, R_{D_{\ell}A}) \rvert^2 \hat{R}_r^{D_{\ell}A} \hat{R}_u^{D_{\ell}A} \\
        + 2 \Re \{ \mathcal{A} (k_1 + k_2, R_{D_{\ell}A}) \left[\mathcal{B}(k_1 + k_2, R_{D_{\ell}A})\right]^{*}\} \hat{R}_r^{D_{\ell}A} \hat{R}_u^{D_{\ell}A}
    \end{split}
    \label{appendix_C_eq_16}
\end{equation}
By substituting Eqs. \eqref{appendix_C_eq_15} and \eqref{appendix_C_eq_16} into Eq. \eqref{appendix_C_eq_10} and after some algebra, one obtains
\begin{equation}
    \begin{split}
        \mathcal{S}_1 = & \sum_{r,u}  \mathcal{U}_{ru} (k_1 + k_2, \vec{R}_{D_{\ell}A}) \mathcal{W}_{ru} (k_{\ell'}, \vec{R}_{D_{1}D_{2}}) \\
        \mathcal{S}_1 = &  3 \lvert \mathcal{A} (k_1 + k_2, R_{D_{\ell}A}) \rvert^2 \lvert \mathcal{A} (k_{\ell'}, R_{D_{1}D_{2}}) \rvert^2 \\
        + &\left\{2 \Re \{ \mathcal{A} (k_{\ell'}, R_{D_{1}D_{2}}) \left[\mathcal{B}(k_{\ell'}, R_{D_{1}D_{2}})\right]^{*}\} + \lvert \mathcal{B} (k_{\ell'}, R_{D_{1}D_{2}}) \rvert^2 \right\} \lvert \mathcal{A} (k_1 + k_2, R_{D_{\ell}A}) \rvert^2 \\
        + &\left\{ 2 \Re \{ \mathcal{A} (k_1 + k_2, R_{D_{\ell}A}) \left[\mathcal{B}(k_1 + k_2, R_{D_{\ell}A})\right]^{*}\} + \lvert \mathcal{B} (k_1 + k_2, R_{D_{\ell}A}) \rvert^2 \right\} \lvert \mathcal{A} (k_{\ell'}, R_{D_{1}D_{2}}) \rvert^2 \\
        + & \left\{2 \Re \{ \mathcal{A} (k_{\ell'}, R_{D_{1}D_{2}}) \left[\mathcal{B}(k_{\ell'}, R_{D_{1}D_{2}})\right]^{*}\} + \lvert \mathcal{B} (k_{\ell'}, R_{D_{1}D_{2}}) \rvert^2 \right\} \\
        \times & \left\{ 2 \Re \{ \mathcal{A} (k_1 + k_2, R_{D_{\ell}A}) \left[\mathcal{B}(k_1 + k_2, R_{D_{\ell}A})\right]^{*}\} + \lvert \mathcal{B} (k_1 + k_2, R_{D_{\ell}A}) \rvert^2 \right\} \left( \hat{R}_{D_{\ell}A} \cdot \hat{R}_{D_{1}D_{2}} \right)^2
    \end{split}
    \label{appendix_C_eq_17}
\end{equation}
where we used $\sum_{r} \hat{R}_r^{D_{\ell}A} \hat{R}_r^{D_{1}D_{2}} = \left(\hat{R}_{D_{\ell}A} \cdot \hat{R}_{D_{1}D_{2}}\right)$.

Finally, the auxiliary function $\mathcal{S}_3$ can be evaluated by separating the summations over repeated indexes
\begin{equation}
    \begin{split}
        \mathcal{S}_3 = \left[\sum_{r,s}  V_{rs} (k_1 + k_2, \vec{R}_{D_{\ell}A}) V_{rs}^{*} (k_1 + k_2, \vec{R}_{D_{\ell}A})\right] \left[\sum_{t,u} V_{tu} (k_{\ell'}, \vec{R}_{D_{1}D_{2}}) V_{tu}^{*} (k_{\ell'}, \vec{R}_{D_{1}D_{2}}) \right]
    \end{split}
    \label{appendix_C_eq_18}
\end{equation}
and, from Eq. \eqref{appendix_C_eq_13}, it is straightforward to obtain
\begin{equation}
    \begin{split}
        \mathcal{S}_3 = &\left[3 \lvert \mathcal{A} (k_1 + k_2, R_{D_{\ell}A}) \rvert^2 + \lvert \mathcal{B} (k_1 + k_2, R_{D_{\ell}A}) \rvert^2 + 2 \Re \{ \mathcal{A} (k_1 + k_2, R_{D_{\ell}A}) \left[\mathcal{B}(k_1 + k_2, R_{D_{\ell}A})\right]^{*}\} \right] \\
       \times & \left[3 \lvert \mathcal{A} (k_{\ell'}, R_{D_{1}D_{2}}) \rvert^2 + \lvert \mathcal{B} (k_{\ell'}, R_{D_{1}D_{2}}) \rvert^2 + 2 \Re \{ \mathcal{A} (k_{\ell'}, R_{D_{1}D_{2}}) \left[\mathcal{B}(k_{\ell'}, R_{D_{1}D_{2}})\right]^{*}\} \right]
    \end{split}
    \label{appendix_C_eq_19}
\end{equation}
These expressions for $\mathcal{S}_1$ and $\mathcal{S}_3$ allow us to evaluate the contributions in Eqs. \eqref{appendix_C_eq_1x}--\eqref{appendix_C_eq_6x} of the collective ICD rate averaged over the molecules' orientations in a compact form. Notice that in Subsection \ref{averaging_over_orientations} we have, for convenience, discussed the far-zone rate, while here the full rate includes the near and intermediate zone are calculated as well.

\section{Explicit expressions of all transition amplitudes derived in Appendix A}
\label{appendix_D}
\allowdisplaybreaks
Here we list all 252 transition amplitudes from Eqs. \eqref{appendix_A_eq_8}--\eqref{appendix_A_eq_10}. We numbered each term for convenience.



\bibliographystyle{apsrev4-2}
\bibliography{bibliography.bib}

\end{document}